\documentclass[prl,aps, twocolumn, nofootinbib,superscriptaddress]{revtex4-1}
\usepackage{amsmath}
\usepackage{amsfonts}
\usepackage{amssymb}
\usepackage{ragged2e}
\usepackage[font=small,labelfont=bf,justification=justified]{caption}

\usepackage{color}

\newcommand{\rev}[1]{\textcolor{blue}{#1}}

  \makeatletter
    \renewcommand\@make@capt@title[2]{%
     \@ifx@empty\float@link{\@firstofone}{\expandafter\href\expandafter{\float@link}}%
      {\textbf{#1}}\@caption@fignum@sep#2\quad}%
    \makeatother
   
\makeatletter 
\renewcommand{\fnum@figure}{\textbf{Figure~\thefigure}}
\makeatother

\renewcommand{\(}{\left(}
\renewcommand{\)}{\right)}

\usepackage{amsmath}
\usepackage{amsfonts}
\usepackage{graphicx}
\usepackage{verbatim} 
\usepackage{subfig}

\newcommand{\mc}{\mathcal}
\newcommand{\real}{\mathbb{R}}

\newcommand{\map}[3]{#1: #2 \rightarrow #3}
\newcommand{\transpose}{\mathsf{T}}

\begin{document}
\bibliographystyle{naturemag}

\title{Developmental increases in white matter network controllability support a growing diversity of brain dynamics}

\author{Evelyn Tang}
\affiliation{Department of Bioengineering, University of Pennsylvania, PA 19104}
\author{Chad Giusti}
\affiliation{Department of Bioengineering, University of Pennsylvania, PA 19104}
\author{Graham Baum}
\affiliation{Brain Behavior Laboratory, Department of Psychiatry, University of Pennsylvania, PA 19104}
\author{Shi Gu}
\affiliation{Department of Bioengineering, University of Pennsylvania, PA 19104}
\author{Eli Pollock}
\affiliation{Department of Physics, University of Pennsylvania, PA 19104}
\author{Ari E. Kahn}
\affiliation{Department of Bioengineering, University of Pennsylvania, PA 19104}
\author{David Roalf}
\affiliation{Brain Behavior Laboratory, Department of Psychiatry, University of Pennsylvania, PA 19104}
\author{Tyler M. Moore}
\affiliation{Brain Behavior Laboratory, Department of Psychiatry, University of Pennsylvania, PA 19104}
\author{Kosha Ruparel}
\affiliation{Brain Behavior Laboratory, Department of Psychiatry, University of Pennsylvania, PA 19104}
\author{Ruben C. Gur}
\affiliation{Brain Behavior Laboratory, Department of Psychiatry, University of Pennsylvania, PA 19104}
\author{Raquel E. Gur}
\affiliation{Brain Behavior Laboratory, Department of Psychiatry, University of Pennsylvania, PA 19104}
\author{Theodore D. Satterthwaite}
\affiliation{Brain Behavior Laboratory, Department of Psychiatry, University of Pennsylvania, PA 19104}
\affiliation{These authors contributed equally.}
\author{Danielle S. Bassett}
\affiliation{Department of Bioengineering, University of Pennsylvania, PA 19104}
\affiliation{Department of Electrical and Systems Engineering, University of Pennsylvania, PA 19104}
\affiliation{These authors contributed equally.}

\date{October, 2016}

\begin{abstract}
As the human brain develops, it increasingly supports coordinated control of neural activity. The mechanism by which white matter evolves to support this coordination is not well understood. We use a network representation of diffusion imaging data from 882 youth ages 8 to 22 to show that white matter connectivity becomes increasingly optimized for a diverse range of predicted dynamics in development. Notably, stable controllers in subcortical areas are negatively related to cognitive performance. Investigating structural mechanisms supporting these changes, we simulate network evolution with a set of growth rules. We find that all brain networks are structured in a manner highly optimized for network control, with distinct control mechanisms predicted in child versus older youth. We demonstrate that our results cannot be simply explained by changes in network modularity. This work reveals a possible mechanism of human brain development that preferentially optimizes dynamic network control over static network architecture.
\end{abstract}

\maketitle


Modern neuroimaging techniques reveal the organization of the brain's white matter microstructure. Collectively, white matter tracts form a large-scale wiring diagram or \textit{connectome} thought to support the brain's diverse dynamics \cite{johansen2010behavioural,hermundstad2014structurally}. Importantly, this architecture changes as children mature into adults \cite{dimartino2014unraveling}, potentially facilitating the emergence of adult cognitive function \cite{baum2016modular}. Despite the intuitive relationship between network structure and brain function \cite{sporns2014contributions}, fundamental mechanistic theories explaining the development of white matter organization and its relationship to emerging cognition in humans have remained elusive. Such a theory would have far-reaching implications for our understanding of normative cognitive development as well as vulnerabilities to neuropsychiatric disorders \cite{Fornito2012}. Indeed, understanding the relations between complex patterns of white matter network reconfiguration and cognitive function could inform interventions to ameliorate cognitive deficits that accompany altered wiring patterns.

Here we investigate how structural connectivity facilitates changes and constrains patterns of dynamics in the developing brain. Drawing on concepts from theoretical physics and engineering, we study two structural predictors of brain dynamics -- controllability \cite{kailath1980linear} and synchronizability \cite{PhysRevLett.80.2109}. We use these two concepts to examine how brains might be optimized for different types of dynamics, and to ask whether individual brains are optimized differently. Controllability is a structural predictor of the ease of switching from one dynamical state to another \cite{pasqualetti2014controllability}, a capability that is critical for traversing a broad state space \cite{gu2015controllability,betzel2016optimally} encompassing a diverse dynamic repertoire \cite{senden2014rich}. Synchronizability is a structural predictor of the ability for regions in the network to support the same temporal dynamical pattern \cite{bassett2006adaptive}, a phenomenon that can facilitate inter-regional communication when implemented locally \cite{fries2015rhythms} but can facilitate pathological seizure-like dynamics when implemented globally \cite{jirsa2014nature,Khambhati20161170}. We hypothesize that white matter networks develop from childhood to adulthood explicitly to maximize controllability and reduce synchronizability.

To test this hypothesis, we examine controllability and synchronizability in structural brain networks derived from diffusion tensor imaging data, which we have represented as weighted adjacency matrices or graphs. We determine the relationship between controllability and synchronizability in a sample of 882 youth from the ages of 8 to 22 \cite{satterthwaite2014neuroimaging}. We demonstrate that networks become optimized for diverse dynamics as children develop, beyond that explainable by changes in the networks' modular structure. Further, we provide supporting evidence for the hypothesis that a balance of controllability across brain regions is required for optimal cognitive function.  

To better understand potential mechanisms of these trajectories, we build a model based on theoretical biology and evolutionary game theory \cite{Avena2014} that describes the observed increase in mean controllability and decrease in synchronizability with age. By exploring changes between networks with similar connection strengths but different connection topologies, we explore the extent to which brain networks are optimized for these architectural features. Then, we define a given subject's capacity to alter its topology towards increasingly diverse dynamics by extracting parameters that govern the speed, extent and fall-off of network optimization. These novel statistics allow us to assess whether children's brains have greater potential for increasing their ability to move from one mental state to another (controllability). Finally, we demonstrate that the evolutionary rule based on controllability and synchronzability is a better fit to the observed empirical data than alternative rules constructed from traditional graph statistics including efficiency \cite{latora2001efficient} and modularity \cite{newman2006modularity}, and including degree and strength. 

\section{Results}

\subsection{Controllability in brain networks}

\begin{figure*}[!htb]
\includegraphics[width=0.98\linewidth]{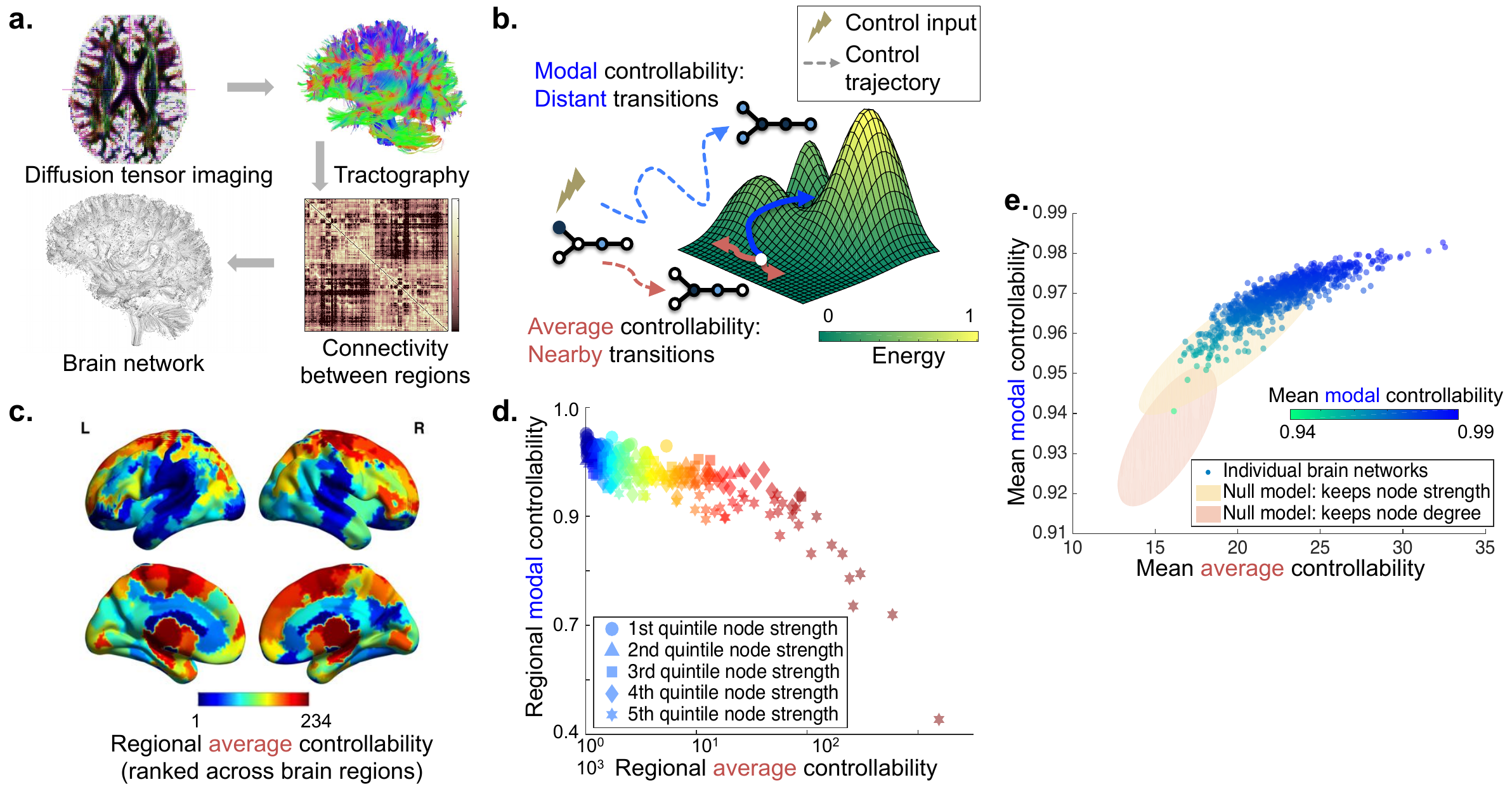}
		\caption{\textbf{Controllability in brain networks. (a)} Diffusion tensor imaging measures the direction of water diffusion in the brain. From this data, white matter streamlines can be reconstructed that connect brain regions in a structural network.  \textbf{(b)} Average controllability: structural support for moving the brain to easy-to-reach states; mean modal controllability: structural support for moving the brain to difficult-to-reach states. \textbf{(c)} Regional average controllability ranked on $N=234$ brain regions of a group-averaged network for visualization purposes. \textbf{(d)} Regions with high average controllability tend to display low modal controllability: $\rho=-0.76$, $df=233$, $p<1\times10^{-16}$; relative node strength is indicated by shape. \textbf{(e)} Controllability measures averaged over all regions in the brain networks of 190 healthy young subjects; each colored circle represents a person. People whose brains display high average controllability also tend to display high modal controllability: $r=0.87$, $df=881$, $p<1\times10^{-16}$. Yellow and red ellipses are the 95\% confidence clouds of network null models in which the edge weights of the brain networks are shuffled to preserve strength or degree, respectively.}\label{fig:control}
\end{figure*}

We begin by asking whether regions of the brain display different predispositions for controllability. To answer this question, we estimate controllability and synchronizability in the structural brain networks of 882 youth from the Philadelphia Neurodevelopmental cohort (Fig.~\ref{fig:control}a; see Methods). We examine two types of controllability, which describe the predicted ability to move the network into different states defined as patterns of regional activity (Fig.~\ref{fig:control}b).  \emph{Average controllability} is a structural phenotype predicted to facilitate small changes in brain state, nearby on an energy landscape. In contrast, \emph{modal controllability} is a structural phenotype predicted to facilitate large changes in brain state, distant on an energy landscape (see Methods). 

To address whether there are related individual differences in types of controllability, we study these metrics in a cohort of 882 youth from ages 8 to 22 (see Methods). In brain networks, nodes with high average controllability tend to be strongly connected, while nodes with strong modal controllability tend to be weakly connected \cite{gu2015controllability}. These nodes are distinct from each other (Fig.~\ref{fig:control}c), indeed regional average controllability is negatively correlated with regional modal controllability (Spearman correlation coefficient $\rho=-0.76$, $df=233$, $p<1\times10^{-16}$; Fig.~\ref{fig:control}d). That is, regions that are theoretically predicted to be good at moving the brain into nearby states are not the same as regions that are theoretically predicted to be good at moving the brain to distant states.

While each brain region may play a different control role, one could ask whether there are related individual differences in types of controllability. To answer this question, we calculate whole-brain average controllability as the mean average controllability value across all brain regions in a single individual, and similarly for whole-brain modal controllability. We find that individuals whose brains display high mean average controllability also display high mean modal controllability (Pearson's correlation coefficient $r=0.87$, $df=881$, $p<1\times10^{-16}$; Fig.~\ref{fig:control}e). This relationship -- which is not characteristic of several common random graph models \cite{wuyan2017benchmarking} -- suggests that brain networks that can support switches among nearby states can also support dynamical transitions among distant states. 

To determine whether these trends in individual variation are expected statistically, we compare our results in the real data with those obtained from corresponding null models. Specifically, we randomly permute the placement of edges weights (i) to preserve strength, or the sum of weights for each node $\sum_jA_{ij}$, or (ii) to preserve degree, or the number of connections for each node $\sum_j|A_{ij}|^0$ (see Methods). We observe that networks in both null models display much lower controllability (both average and modal) than the true data (Fig.~\ref{fig:control}e), particularly when only degree is preserved. For average controllability, the true data has mean and standard deviation of $22.6\pm2.5$, while the null models show (i) $19.5\pm2.2$ and (ii) $15.7\pm 1.2$ respectively. For modal controllability, the true data has mean and standard deviation of $ 0.969\pm 0.006$, while the null models show (i) $0.958\pm0.008$ and (ii) $0.934\pm 0.007$ respectively. $T$-tests to compare the true average controllability data with the null models show that they are significantly different with $p<1\times10^{-16}$ for both null models, with effect sizes measured by Cohen's $d$ of (i) $d$=1.27 and (ii) $d$=3.43. Similar results were obtained when comparing the true modal controllability data with both null models, which showed a significant difference with $p<1\times10^{-16}$ and effect sizes of (i) $d$=1.68 and (ii) $d$=5.44 respectively. These clear differences are striking considering the fact that both null models still inherit many traits from the original networks, including the number of nodes and the weight distributions. This suggests that brain networks are particularly optimized for high controllability to both nearby and distant states, and that this optimization differs across individuals. In particular, we examine the regression slope between mean average controllability and mean modal controllability in these three network ensembles using an analysis of covariance test to find that (i) random, brain-like networks with preserved strength have a difference in regression slope with empirical brain networks of  $t=14.5$,  $df=1760$, $p<1\times10^{-16}$ (ii) random, brain-like networks with preserved degree have a difference in regression slope with empirical brain networks of $t=16.7$, $df=1760$, $p<1\times10^{-16}$.

\begin{figure*}[tb]
\includegraphics[width=0.65\linewidth]{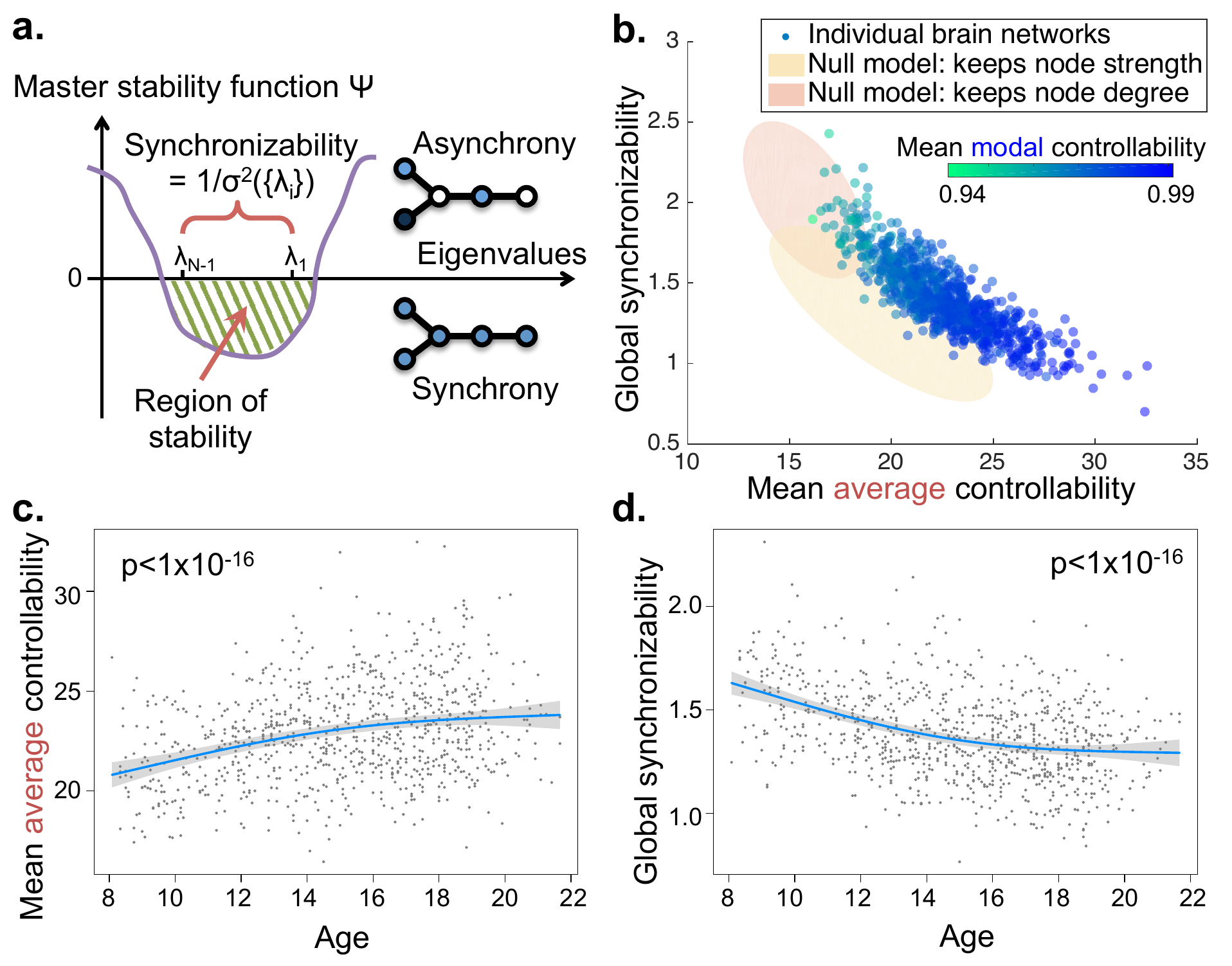}
\caption{ \textbf{Synchronizability and changes across development. (a)} A synchronous state is operationalized as a state in which all nodes have the same activity magnitude. Such a state is stable when the master stability function is negative for all positive eigenvalues of the graph Laplacian (see Methods). We use the inverse spread of the Laplacian eigenvalues $1/\sigma^2(\{\lambda_{i}\})$ as a measure of global synchronizability. \textbf{(b)} Global synchronizability is anti-correlated with both average controllability and modal controllability (color of circles). Yellow and red ellipses are the 95\% confidence clouds of the node-preserving and strength-preserving null models. \textbf{(c)} Mean average controllability significantly increases with age: Pearson's correlation coefficient $r=0.28$, $df=881$, $p<1\times10^{-16}$. \textbf{(d)} Global synchronizability significantly decreases with age: $r=-0.37$, $df=881$, $p<1\times10^{-16}$. The fits in panels \emph{(c, d)} all control for brain volume, head motion, sex, and handedness. Blue lines show best non-linear fit under a general additive model (see Methods); gray envelope denotes 95\% confidence interval.}\label{fig:spatial}
\end{figure*}

\subsection{Synchronizability and changes across development}

While controllability predicts the ability of a network to change between states, synchronizability predicts the ability of a network to persist in a single (synchronous) state. Mathematically, this property of a complex system can be studied using the master stability function \cite{PhysRevLett.80.2109,barahona2002synchronization}. Specifically, stability under perturbations exists when this function is negative for all positive eigenvalues of the graph Laplacian $\{\lambda_{i}\}, i=1, ... ,(N-1)$, or -- put another way -- when all $\{\lambda_{i}\}$ fall within the region of stability (Fig.~\ref{fig:spatial}a). A larger spread of Laplacian eigenvalues will make the system more difficult to synchronize, and therefore an intuitive measure of global synchronizability is the inverse variance $1/\sigma^2{(\{\lambda_{i}\})}$ \cite{Nishikawa08062010} (see Methods for details and an illustration of this dominant contribution from the eigenspectrum variation in Supplementary Fig. 1).

Using this theoretical scaffold, we observe that brain networks that are more synchronizable tend to display lower average controllability (Pearson's correlation coefficient $r=-0.85$, $df=881$, $p<1\times10^{-5}$; Fig.~\ref{fig:spatial}b) as well as lower modal controllability  ($r=-0.82$, $df=881$, $p<1\times10^{-5}$). While no known relationship between synchronizability and controllability exists, the correlation is intuitive in that it suggests that individuals who are theoretically predicted to more easily transition into a variety of dynamical states are less susceptible to having many regions locked in synchrony. 
 
We observe that average controllability increases as children age (Pearson correlation coefficient $r=0.28$, $df=881$, $p<1\times 10^{-16}$; Fig.~\ref{fig:spatial}c), as does modal controllability ($r=0.22$, $df=881$, $p=3.5\times10^{-11}$, controlled for brain volume, head motion, sex and handedness). Moreover, we observe that synchronizability decreases as children age ($r=-0.37$, $df=881$, $p<1\times10^{-16}$; Fig.~\ref{fig:spatial}d). These developmental arcs suggest that as the brain matures, its network architecture supports a larger range of dynamics (from nearby to distant states) -- more diverse dynamics -- and is less able to support globally synchronized states. It is natural to ask whether these observations can be simply explained by changes in network modularity that accompany development \cite{baum2016modular}. We show in a later section of the Results, entitled ``Controlling for modularity'', that these results still hold after regressing out network modularity from the variables of interest.

\subsection{Super\rev{-}controllers and cognition}

\begin{figure*}[!htb]
\includegraphics[width=0.65\linewidth]{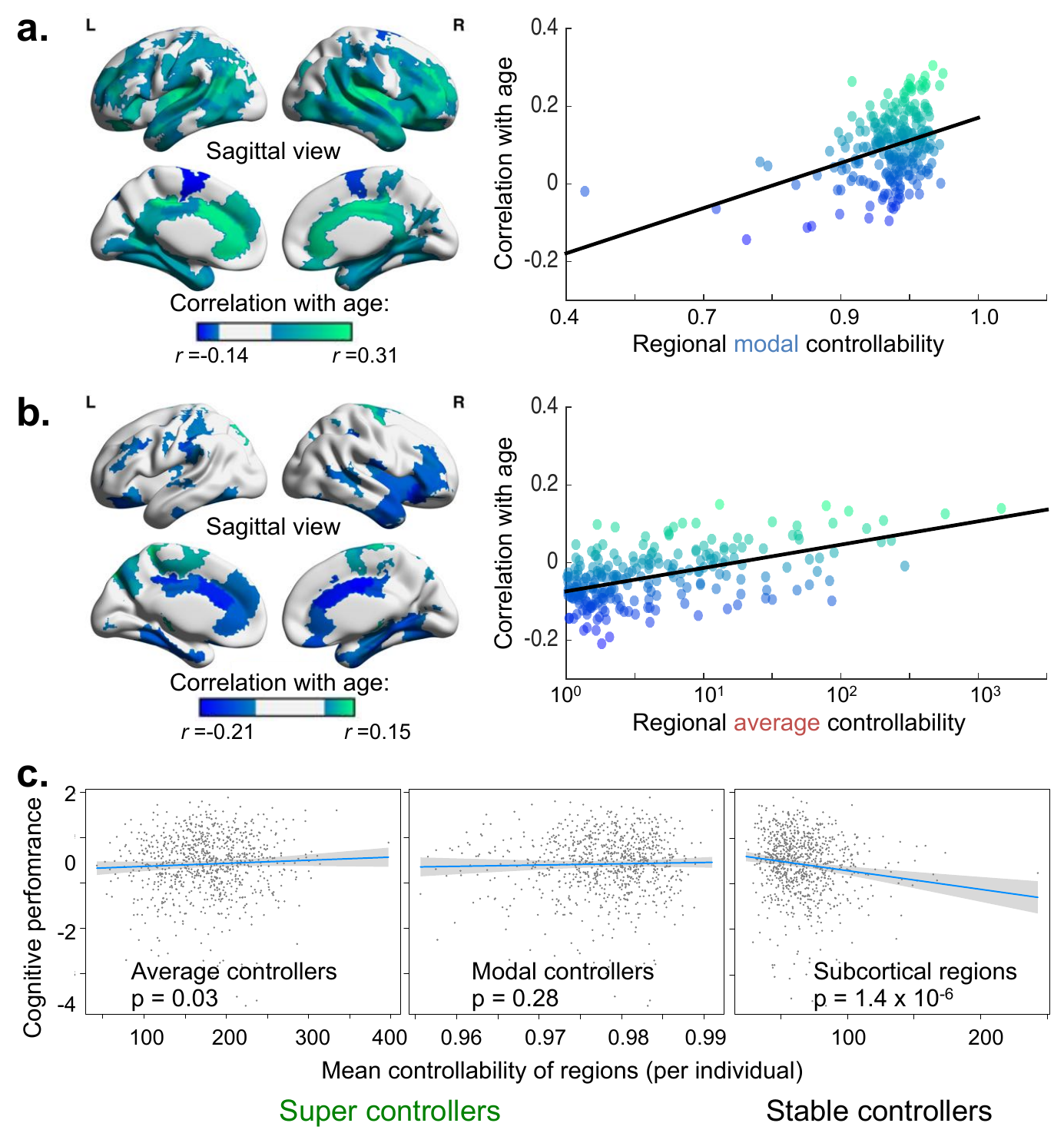}
\caption{\textbf{Regional specialization with age and its impact on cognition. (a)} \emph{(Left)} Regions of significantly increasing modal controllability with age (green) and significantly decreasing modal controllability with age (dark blue). \emph{(Right)} The green regions tend to be stronger modal controllers (`super-controllers'), as seen by the positive slope between the age correlation and regional modal controllability values. \textbf{(b)} \emph{(Left)} Regions of significantly increasing average controllability with age (green) and significantly decreasing average controllability with age (dark blue). \emph{(Right)} The green regions tend to be stronger average controllers (`super-controllers'), as seen by the positive slope between the age correlation and regional average controllability values. \textbf{(c)} \emph{(Left)} Super average controllers (green regions that significantly increase in controllability with age and tend to have higher average controllability) show little relation with cognitive performance (age-normed). The blue line denotes the best linear fit and the grey envelope denotes the 95\% confidence interval. \emph{(Center)} Super modal controllers also show little relation with cognitive performance.  \emph{(Right)} The regions that are most stable in controllability over development -- subcortical regions -- show a significant negative correlation between their average controllability and cognitive performance. These results suggest that the relative strength of controllers in subcortical \emph{versus} cortical regions is critical for understanding individual differences in overall cognitive function; i.e. a shift in control away from cortical regions may be detrimental to higher-order cognition. The fits in panels \textbf{(c)} all control for age, brain volume, head motion, sex, and handedness.} \label{fig:regionalage}
\end{figure*}

Given the global trends of increasing controllability and decreasing synchronizability with age, it is worth asking whether specific regions of the brain are driving these changes, or whether all regions contribute equally. Surprisingly, we observe that the regions that display the most controllability also show the greatest developmental increase in control, while regions with lower controllability decrease further with age (Fig.~\ref{fig:regionalage}a--b; figures are averaged over 882 subjects). Regions that increase the most in average controllability show a Spearman correlation with their average controllability value of $\rho=0.48$, $p<1\times10^{-16}$, while regions that increase the most in modal controllability show a correlation with their modal controllability value of $\rho=0.33$, $p<4\times10^{-7}$. We refer to these strong controllers that increase in controllability with age as `super-controllers', whose putative role in the network lies in the differentiation of brain structure necessary to support the wider variety of dynamics that accompanies normative maturation. An average super-controller is therefore a region with high average controllability, and whose average controllability increases with age; a modal super-controller is therefore a region with high modal controllability, and whose modal controllability increases with age. As these significant age associations are found widely across regions in the brain  (see Fig.~\ref{fig:regionalage}a--b and Supplementary Tables I-II for all regions implicated), this suggests that our results are not driven by, for instance, the contribution of the maturation of a single tract. We verify this by examining the network edges that show significant changes with age across all 882 subjects in our youth sample -- i.e. edges that are significantly correlated with age and pass the Bonferroni test for multiple comparisons at $p<0.05$, while controlling for sex, handedness, brain volume and head motion. We find that such edges (95 out of 24027 non-zero (unique) edges in the 234-region network) are distributed very broadly throughout the brain, connecting 43\% of the 234 brain regions. This evidence of contributions from the maturation of many tracts instead of a single tract, supports the need for larger scale descriptors such as network controllability, with which to build models.

Regions of high modal controllability (which tend to have low degree, see section ``Pareto optimization with other metrics'') effect energetically distant state transitions and -- in healthy adult subjects -- are disproportionately located in cognitive control regions, while regions of high average controllability (which tend to have high degree structure, see section ``Pareto optimization with other metrics'') effect energetically nearby state transitions and -- in healthy adult subjects -- are disproportionately located in several brain systems including the default mode \cite{gu2015controllability}. As we test statistically in the previous paragraph, we find a global strengthening of these super-controllers, consistent with an increasing specialization of function that accompanies development and experience. Regionally, we observe that some controllers in prefrontal cortex and ACC increase in their modal controllability with age and decrease in their average controllability with age, as we show statistically in the previous paragraph, potentially suggesting a narrowing or specialization of their preferred control roles with development.

Speculatively, it may be the case that the emergence of super-controllers over the course of development could explain differences in cognitive function. Alternatively, it is possible that these super-controllers are unstable points in the network undergoing massive re-organization with age, and therefore that optimal predictors of individual differences in cognitive function (above and beyond that expected by age) will instead be found in the regions that remain stable in their controllability over development. To test this pair of conflicting hypotheses, we examine the relationship between cognitive performance on a battery of tasks and individual differences in controllability, separately averaged over (i) average super-controllers (Fig.~\ref{fig:regionalage}c, left), (ii) modal super-controllers (Fig.~\ref{fig:regionalage}c, center), and (iii) stable-controllers, or those regions whose controllability did not significantly change with age (Fig.~\ref{fig:regionalage}c, right). While controlling for the effects of age, we observe that individuals with higher cognitive performance (see Methods) also display weaker stable-controllers, largely located in subcortical areas (Spearman correlation coefficient between cognitive performance and mean average controllability of stable-controllers $\rho=-0.16$, $df=879$, $p=1.4\times10^{-6}$). Note that the use of Spearman correlations eliminates the dependence of these results on any outliers, and only (iii) passes a false discovery rate correction for multiple comparisons across the three tests (the other $p$-values are (i) $p=0.03$ and (ii) $p=0.28$ respectively). These results suggest that the relative strength of controllers in subcortical \emph{versus} cortical regions is critical for understanding individual differences in overall cognitive function, i.e. a shift in control away from cortical regions may be detrimental to higher-order cognition.
\begin{figure*}[tb]
\includegraphics[width=0.65\linewidth]{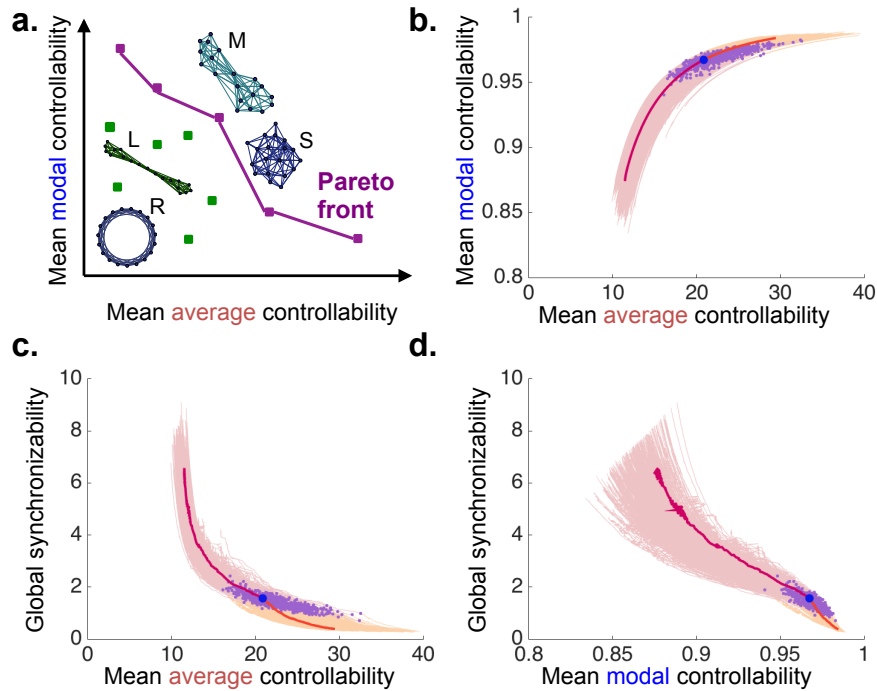}
\caption{\textbf{Brain networks are optimized for diverse dynamics. (a)} Pareto optimization explores a family of networks with different topologies and hence varying mean controllability and synchronizability (a few toy models illustrate this including a ring lattice R, regular lattice L, modular network M, and small-world network S). Pareto optimal networks (purple dots) are the networks where these properties are most efficiently distributed, i.e., it is impossible to increase one property without decreasing another property -- unlike in the non-optimal networks (green dots). The boundary connecting the Pareto-optimal networks forms the Pareto front (purple line). \textbf{(b, c, d)} Beginning from an empirically measured brain network (purple dots), we swap edges to modify the topology and test if the modified network advances the Pareto front. This procedure charts a course of network evolution characterized by increasingly optimal features: here we increase mean average controllability and mean modal controllability, and decrease global synchronizability, in 1500 edge swaps (yellow curves). For comparison, we also evolved the network in the opposite direction (to decrease controllability and increase synchronzability, pink curves). The trajectory for one subject (blue dot) is highlighted (orange and red). See Methods for evidence of convergence of controllability metrics in the forward direction after 1500 edge swaps.} \label{fig:paretocurves}
\end{figure*}

\subsection{A network growth model for modeling development}

Thus far, we have demonstrated that network controllability and synchronizability appear to follow a characteristic curve (Figs.~1e \& 2b), change significantly with age (Fig.~2c-d), and correlate with individual differences in cognition (Fig.~3c). Yet, none of these observations constitute a mechanistic theory. However, the first 2 of these observations does suggest one: brain networks develop \emph{explicitly} to maximize controllability while limiting synchronizability. To test this hypothesis, we use an evolutionary algorithm to chart a course for network evolution in the 3-dimensional space of these features (average controllability, modal controllability, and synchronizability). We employ an optimization method developed in economics and game theory, Pareto optimization, which has recently been adapted to explore brain network topologies (morphospace) \cite{Avena2014}. Beginning with a brain network obtained from the original data, an existing edge in the network is randomly chosen for rewiring, to take the place of an edge that did not previously exist. The controllability and synchronizability metrics are calculated for that new network and if the new network is found to advance the Pareto front (see Fig.~\ref{fig:paretocurves}a), the rewiring is retained; if not the rewiring is dismissed. This process is repeated to chart a course by which networks increase controllability and decrease synchronizability, while maintaining the same edge weight distribution and mean degree. To provide contrast in the opposite direction, we evolve the subject's network both forward in developmental time (increasing control and decreasing synchronizability), and backward in developmental time (decreasing control and increasing synchronizability).

Critically, we observe that the simulated evolutionary trajectory that optimizes controllability and minimizes synchronizability is a constrained path that tracks the human brain data points well (mean and variance of the distance from the data points to the average predicted curve is $0.0049\pm0.0376$). These results support the hypothesis that a mechanism of human brain development is the reconfiguration of white matter connectivity to increase the human's ability to flexibly move between diverse brain states \cite{chai2017evolution}. In addition to this fundamental and more general insight, we also make several specific observations, which we detail thoroughly in the next section.

\subsection{Brain networks are near optimal for controllability}

Here we investigate the trajectories traced out by evolving networks upon optimizing for controllability and synchronizability metrics. First, we demonstrate that brain networks are well optimized for high controllability and low synchronizability by comparing distances travelled in the forward and backward directions. Second, we compare the evolved metrics with the data to show brain networks can reach near-optimal values of controllability but seem to saturate at a finite level of synchronizability.

Our Pareto-optimization algorithm runs for 1500 edge steps in both the forward (optimizing for high average and modal controllabilities, and low synchronizability) and backward (optimizing for low average and modal controllabilities and high synchronizability) directions. An estimate of the discrete distance traveled in the forward direction is 
\begin{equation}
d_f=\sqrt{\(\frac{x_f}{x_0}\)^2+\(\frac{y_f}{y_0}\)^2+\(\frac{z_f}{z_0}\)^2}
\end{equation}
where $x$ is mean average controllability, $y$ is mean modal controllability and $z$ is synchronizability. This is a dimensionless distance, normalized by the total distance travelled, i.e. $x_0=x_f-x_b$, where $x_f$ and $x_b$ are changes in the forward and backward directions, respectively. Similarly, we can write a similar expression for the dimensionless distance in the backward direction by replacing $f\to b$ in the above expression. The ratio of distance traveled forward to backward is then $d_f/d_b$.

Examining first the average controllability \emph{versus} modal controllability plane (Fig.~4b and setting $z=0$ in the expression above), we find that this ratio is 0.52, so it is almost twice as easy to decrement the controllability values than to increase them. Including synchronizability as well in the full three-dimensional space (Figs.~4c--d), we find that this ratio is 0.46, indicating that it is also markedly easier to increase synchronizability than to decrease it. These results indicate that within the space of networks with the same edge distribution, brain networks have topologies that are well optimized for high controllability and low synchronizability.

\begin{figure}[tb]
\includegraphics[width=0.95 \linewidth]{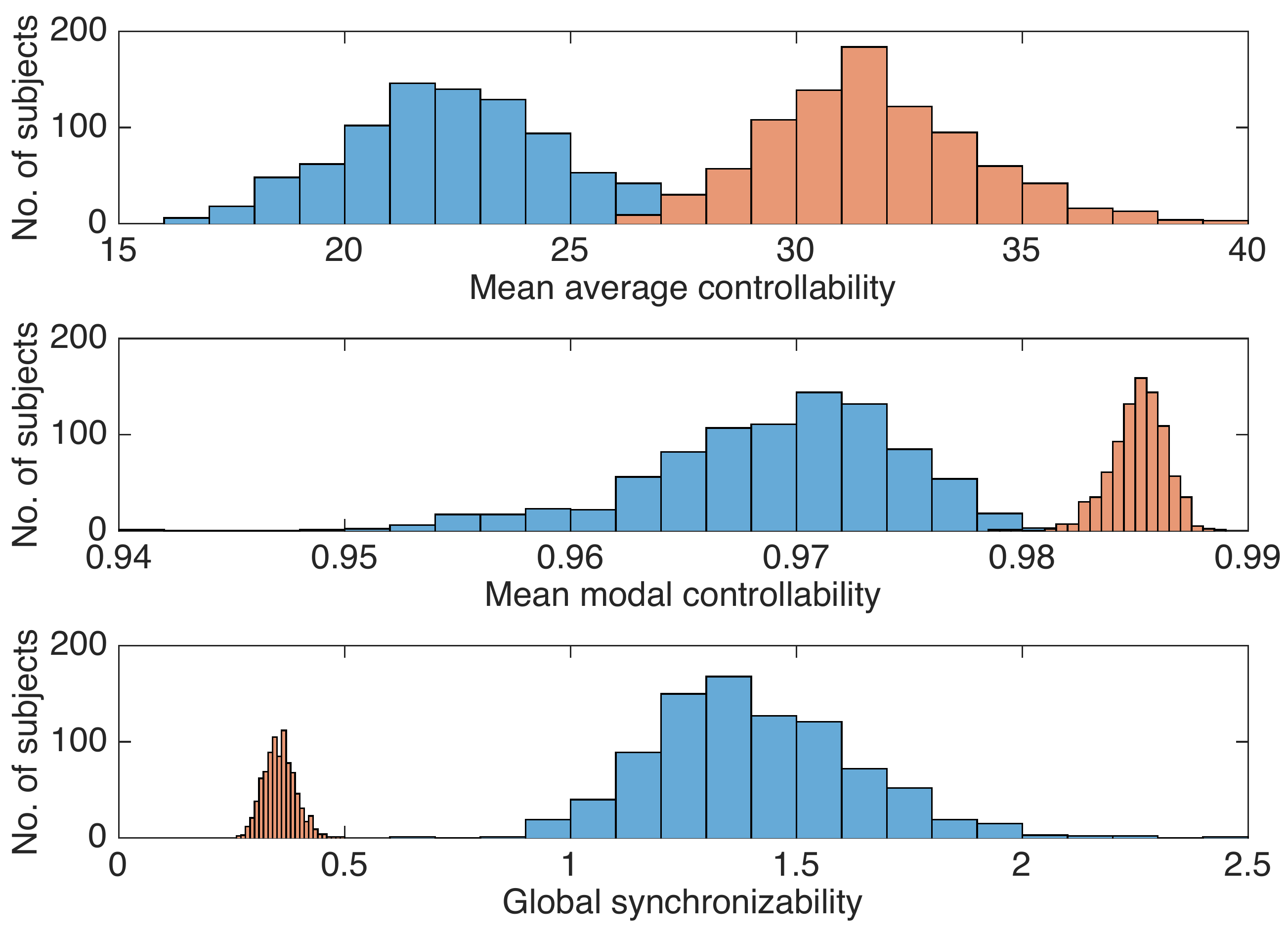}
\caption{\textbf{Brain networks show near-optimal control but finite synchronizability.} The original brain networks (blue) and final evolved networks (orange) show overlap between their controllability values (first and second plots), however there is no overlap between the synchronizability values of these two groups (third plot) --- suggesting that brain networks display near optimal control but retain a finite level of synchronizability.} \label{fig:nearoptimal}
\end{figure}

The final evolved values for controllability (31.7 for average controllability and 0.985 for modal controllability) are more like actual values shown by brain networks (maximum average controllability is 32.6 and maximum modal controllability is 0.983) than are the final evolved values for synchronizability, see Fig.~\ref{fig:nearoptimal}. This suggests that brain networks have near-optimal controllability, but do not fully limit synchronizability, perhaps because some finite amount of synchronization is needed for dynamical coordination and cognition.

\subsection{Pareto optimization with other metrics}

In this section, we provide comparisons with related network metrics such as maximum and minimum weighted degree (while preserving mean weighted degree), to demonstrate the specificity of controllability metrics. As controllability metrics describe the propagation of dynamics in the network, they dramatically constrain evolutionary trajectories much more than simply increasing the maximum or minimum weighted node degrees. We also find that optimization using other relevant network metrics such as global efficiency \cite{latora2001efficient} and network modularity \cite{newman2006modularity} displays far less structure as compared to optimizing for controllability and synchronizability (see later subsection in the Results entitled ``Controlling for modularity'').
In comparison to the complete set of available models, we demonstrate that brain networks are structured in a manner best described as highly optimized for the control of diverse neural dynamics.

\begin{figure}[tb]
\includegraphics[width=\linewidth]{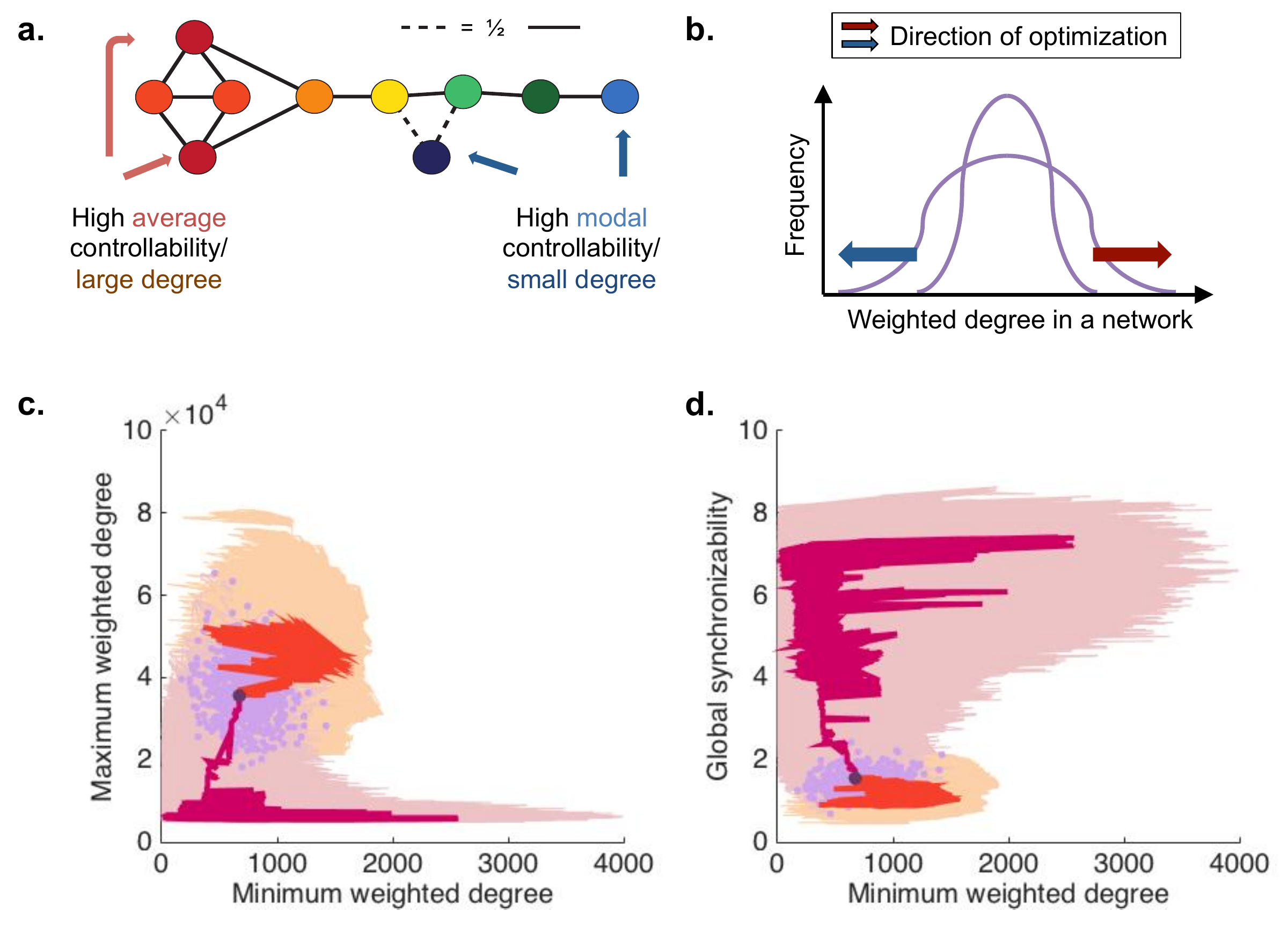}
\caption{\textbf{Specificity of controllability \textit{versus} degree. a.} As average (modal) controllability of each node is closely tied to high (low) weighted degree of that node \cite{gu2015controllability}, we repeat our optimization for maximum and minimum weighted degree in the network. \textbf{b.} While the edge weight distribution of the network remains the same, edge swaps can alter the total degree of each node to increase the minimum (blue arrow) or maximum (red arrow) degree of the nodes. \textbf{c, d.} The maximum and minimum weighted degree of each subject's brain network are plotted as purple dots (\textbf{c.}) and similarly for minimum weighted degree and synchronizability (\textbf{d.}) --- we see little structure or discernible relationship between individuals. We also plot representative optimization trajectories for each subject in the forward direction (yellow) in the cross section of increasing maximum weighted degree and decreasing minimum weighted degree (\textbf{c.}) and decreasing minimum weighted degree and synchronizabililty (\textbf{d.}); as well as trajectories in the opposite direction (pink). The trajectory for a single representative individual is highlighted, for both forward (red) and backward (dark red) directions. We observe that this example trajectory takes a meandering path through the plane, displaying little structure.}\label{fig:strengthcont}
\end{figure}

The weighted degree of each node has a strong overlap with the controllability of that node, see Fig.~\ref{fig:strengthcont}a. Here, we verify that modifying the degree structure of each node -- in a way similar to changes wrought by optimizing for controllability -- does not simply recapitulate the results given by optimizing for controllability. Weighted degree is the sum of all the edges connected to that node; given the adjacency matrix $A_{ij}$ it is $\sum_jA_{ij}$ for node $i$. The average controllability of a node has a strong positive correlation with ranked weighted degree and the modal controllability of a node has a strong negative correlation with ranked weighted degree \cite{gu2015controllability}.

Hence, a matrix that simultaneously increases mean average controllability and mean modal controllability could simply be a matrix that increases its largest and smallest weighted degree -- thereby stretching out the degree distribution (Fig.~\ref{fig:strengthcont}b). While our edge swapping procedure does not alter the edge weight distribution or mean weighted degree of the network, the total degree of each node can be altered to increase the minimum or maximum weighted degree respectively. We repeat our simulations now optimizing for an increase in the maximum weighted degree, and a decrease in the minimum weighted degree and global synchronizability. If controllability is merely a proxy for weighted degree, then this should give similar results to the simulations that optimize for increased controllability and decreasing synchronizability.

First, we observe that plotting the raw data according to maximum and minimum weighted degree (purple dots in Fig.~\ref{fig:strengthcont}c) reveals very little structure, unlike the clean developmental arc seen in Fig.~\ref{fig:control}e. This is also true for the plot of minimum weighted degree and global synchronizability (purple dots in Fig.~\ref{fig:strengthcont}d), where this is little discernible relationship, unlike the clean developmental arc shown in Fig.~2b. Second, instead of the constrained curves we see in the forward trajectories of Fig.~\ref{fig:paretocurves} that mimic the developmental arc, now the paths simply move in a noisy manner across the plane (Fig.~\ref{fig:strengthcont}c, d: forward trajectories in yellow and backward trajectories in pink). In the highlighted trajectory for a single individual (orange and red) we see that this curve zigzags across the plane. Moreover, trajectories from separate simulations do not overlap with one another as do trajectories from separate simulations that optimize for controllability metrics.

Together, these results demonstrate that controllability metrics are far more constrained than weighted degree, or rather, high weighted degree appears to be necessary but not sufficient for average controllability, and similarly for low weighted degree and modal controllability. The derivation of controllability metrics comes from a specific dynamical model that utilizes network connectivity for the propagation of dynamics, and is far more constraining than simply having many large driver nodes or many poorly connected nodes.

\subsection{Steeper trajectories in children \textit{versus} older youth}

\begin{figure*}[tb]
\includegraphics[width=0.65\linewidth]{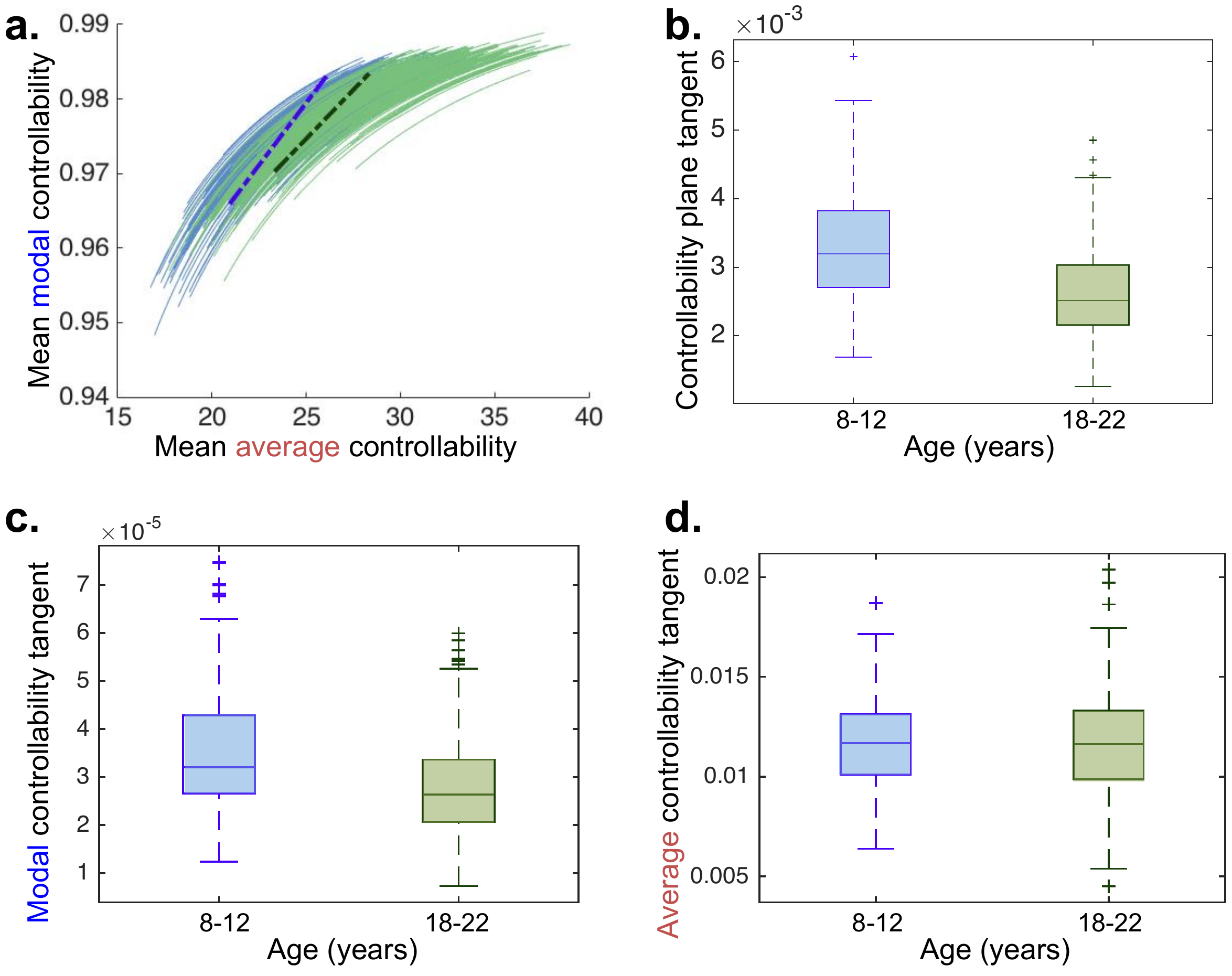}
\caption{\textbf{Steeper trajectories in children \textit{versus} youth ages 18 to 22.  (a)} We compare two cohorts of different ages, 170 children from ages 8 to 12 years (blue) and 190 youth from ages 18 to 22 years (green). Based on their forward Pareto-optimization trajectories, we fit exponential curves $y=a+b \exp(cx)$ for each subject, where $x$ is the mean average controllability and $y$ is the mean modal controllability, to obtain the curve tangent at the position of the brain network. The mean tangents of both groups are shown as dotted lines: the children's in blue and the older youth's in green. \textbf{(b)} The distribution of tangents for both groups shows that children have a 27\% steeper slope in their optimization curves as compared to the youth ages 18 to 22; non-parametric permutation test $p<0.001$. \textbf{(c)} This difference arises from the evolutionary curve in modal controllability, as can be seen by a separate examination of how modal controllability changes with each re-wiring step (group-difference in tangent: $p<0.001$) -- suggesting that children have a greater potential for increasing their ability to make distant or difficult changes in mental state than older youth. \textbf{(d)} In contrast, there is little group-difference in the change of average controllability with each re-wiring step ($p=0.47$), suggesting the potential to increase nearby mental switches remains constant over development.}  \label{fig:devtangents}
\end{figure*}

In the previous section, we provided important evidence to support a mechanistic theory that implicates network reconfiguration towards optimal controllability as a fundamental driver of neurodevelopment. Next, we turn from the global assessment to the individual, and study the charted evolutionary trajectories of each subject to ask whether that simulated trajectory harbors important information regarding the subject's age and predictions regarding the subject's abilities. We begin by studying the capacity for a brain network to adapt by estimating the tangent of the evolutionary trajectory. For each simulated trajectory, we first fit the exponential form $y=a+b \exp(cx)$ to the average and modal controllability, and estimate the tangent of the curve at the position of the actual brain network (Fig.~\ref{fig:devtangents}a). By comparing the group difference in these tangents between children from 8 to 12 years ($n=170$) and youth from 18 to 22 years ($n=190$) using a non-parametric permutation test, we find that children display larger tangents with $p<0.001$, see Fig.~\ref{fig:devtangents}b -- and hence steeper evolutionary curves. These results suggest that children's brain networks have a greater capacity for network evolution than the brain networks of older youth, as we show through investigating the separate contributions to these changes from modal and average controllability. This revealed that the group difference in steepness of the evolutionary curve is driven much more by the change in only modal controllability as a function of rewiring step (Fig.~\ref{fig:devtangents}c; group-difference in tangent: non-parametric permutation test $p<0.001$) than by the change in only average controllability as a function of rewiring step (Fig.~\ref{fig:devtangents}d; $p=0.47$). These results suggest that children have a greater potential to increase their ability to make distant or difficult changes in mental state more than youth ages 18 to 22, whereas the potential to increase their nearby mental switches remains constant over development.

\subsection{Controlling for modularity}
Given that structural brain networks are modular \cite{sporns2016modular} and that modularity changes with age \cite{baum2016modular}, one might ask if modularity is related to controllability or synchronizability, or if changes in modularity can explain changes in these metrics with age. In this section, we describe a set of analyses that demonstrate that (i) modularity and controllability do not have a one-to-one correspondence, and in fact show very different dependencies across different graphs, (ii) our results hold after controlling for the modularity quality index $Q$, and (iii) Pareto-optimization trajectories based on modularity and efficiency do not recapitulate the trends observed in the empirical data. Here, we provide brief summaries of each of these tests, and we refer the reader to the Supplementary Results subsection ``Controlling for modularity'' and Supplementary Figs. 5-6 for additional details.

First, we note that to our knowledge, there are no analytical results relating modularity to controllability or synchronizability thus far. Hence we ask whether modularity and controllability can be observed to be numerically correlated with one another over instances in a graph ensemble. Even without explicit mathematical dependence, numerical correlations between two variables can still occur -- in the simplest case -- if a third variable is driving changes in both. We demonstrate that modularity has no instrinsic relationship with controllability by investigating various families of graph models, each with two different types of edge weight distributions. The graph models chosen include the weighted random graph, the ring lattice graph, the Watts-Strogatz small-world graph, three examples of a modular graph (with 2, 4 and 8 modules), the random geometric graph, and the Barabasi-Albert preferential attachment graph. Edge weights were drawn either from a Gaussian distribution or from an empirical distribution of streamline counts.  For each of these families of networks, ensembles of 100 networks each were generated. Scatterplots of the modularity quality index $Q$ and mean average controllability are given for each of these 2 (edge weights) x 8 (graph model) types of network ensembles Supplementary Figs. 5-6 and a summary table of the Spearman correlations are given in Table \ref{tab:spearman2}. From these data, it is evident that the correlations can vary across graph types from strongly positive ($\rho=0.14$, $p=0.16$), to strongly negative ($\rho=-0.75$, $p<1\times10^{-16}$), to no relation at all ($\rho=-0.03$, $p=0.80$). These simulations clearly demonstrate that there is no \emph{a priori} relation between the dynamical predictors of controllability and synchronizability, and the heuristic of modularity.

\begin{figure*}[tb]
\includegraphics[width=0.75\linewidth]{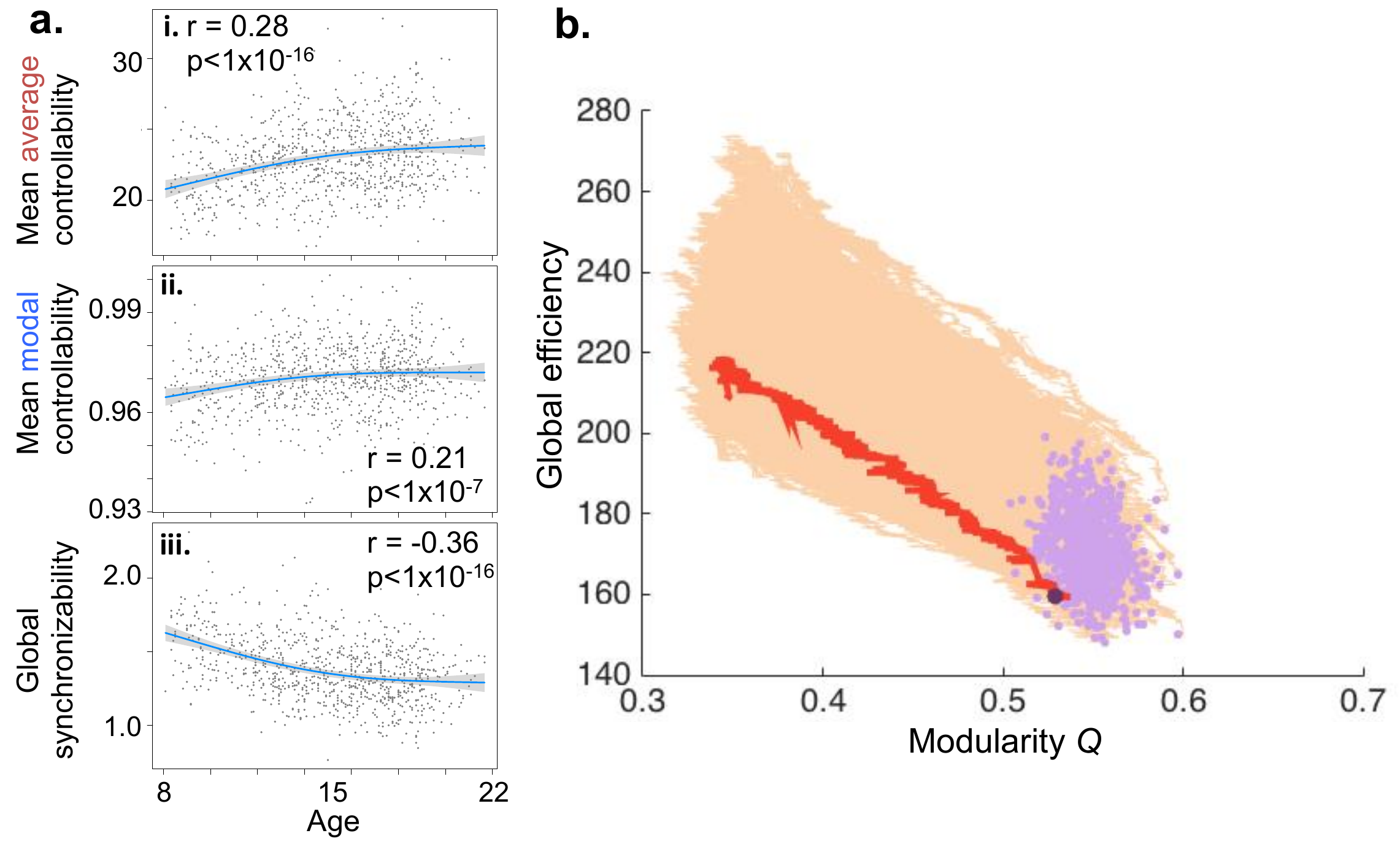}
\caption{\textbf{Ruling out a dependence on modular structure in our results. (a)} Linear fits when including the modularity $Q$ index as a covariate, for plots of \textbf{(i)} mean average controllability, \textbf{(ii)} mean modal controllability, and \textbf{(iii)} global synchronizability respectively -- their relationships with age are barely changed.  \textbf{(b)} The Pareto-optimization of brain networks for the metrics of modularity $Q$ and global efficiency, does not display a similar functional form to the empirical data under these two metrics (the light purple dots). The simulated trajectories in the forward direction of these two metrics are shown in yellow. For illustration purposes, the large, dark purple dot and red trajectory indicate the result for a single individual.}  \label{fig:modularity}
\end{figure*}

We next ask whether individual differences in modularity can explain changes in controllability or synchronizability that occur with age. To address this question, we calculate the modularity metric $Q$ by running 100 iterations of a Louvain-like locally greedy algorithm to maximize the modularity quality function with $\gamma=1$ for each subject's structural adjacency matrix \cite{bassett2013robust}. For each subject we then obtain a consensus partition \cite{bassett2013robust} and the consensus $Q$ value. By recalculating the relationships between synchronizability and controllability while regressing out the effects of modularity, we find that our results remain robust -- mean average controllability remains strongly correlated with mean modal controllability ($r=0.87$, $df=881$, $p<1\times10^{-5}$), and both quantities are negatively correlated with synchronizability:  $r=-0.84$, $df=881$, $p<1\times10^{-5}$ for mean average controllability and  $r=-0.81$, $df=881$, $p<1\times10^{-5}$ for mean modal controllability, respectively. In addition, mean average controllability and mean modal controllability remain positively correlated with age ($r=0.28$, $df=881$, $p<1\times10^{-5}$ and $r=0.21$, $df=881$, $p<1\times10^{-5}$, respectively), while synchronizability is negatively correlated with age ($r=-0.36$, $df=881$, $p<1\times10^{-5}$), with very little change in the magnitude of the results; see Fig. \ref{fig:modularity}a. 

Finally, we repeat the Pareto optimization in two dimensions to optimize for modularity as well as for global efficiency, which has been suggested as an important factor in brain network evolution in prior studies. For instance, the Pareto-optimization of brain networks was first conducted during the investigation of the structural evolution of various biological networks using the metric of global efficiency \cite{Avena2014}. Separately, brain networks have demonstrated changes in modularity with age \cite{Hagmann02112010}. Hence we repeat the Pareto optimization in two dimensions to optimize for global efficiency and modularity, in order to provide a dialogue with previous work in the literature. To calculate global efficiency \cite{latora2001efficient}, the function \textit{efficiency\_wei} from the Brain Connectivity Toolbox was used, while a generalized Louvain-like \cite{blondel2008} locally greedy community detection method \cite{Mucha876} was used to optimize modularity \cite{newman2006modularity}, by comparison with a standard Newman-Girvan null model \cite{girvan2002community}.

We optimize for these two quantities in a two-dimensional space for 1500 iterations in the forward direction, where we find that separate trajectories for the same subject have a strong overlap (the summed square difference between a single trajectory for that subject and that subject’s average trajectory, is only 6.4\% of the this difference across different subjects). These optimization trajectories (fairly linear) do not display a similar functional form to the empirical brain networks under these two metrics (which do not have much structure), see Fig. \ref{fig:modularity}b. This difference in the functional form observed in the true and simulated data drawn from a Pareto-optimization of modularity and global efficiency stands in contrast to the similarity in functional form observed when the simulated data is drawn from a Pareto-optimization of controllability and synchronizability -- where these latter metrics on the empirical brain networks can be fit to similar exponential forms as the Pareto-optimal trajectories (Section ``Steeper trajectories in children \textit{versus} older youth''). In contrast, these data demonstrate that modularity and global efficiency are not parsimonious candidates for network-level mechanisms of structural rewiring in brain networks over development.

Taken together, these results suggest that modularity does not provide a compelling explanation for the observed, age-related relationship between controllability and synchronizability in brain networks.

\section{Discussion}

We address the fundamental question of how the architecture of the brain supports the emergence of cognitive abilities in humans. To do so, we drawon the computational tools and conceptual frameworks of theoretical physics and engineering to study two complementary predictors of brain dynamics -- controllability and synchronizability -- built from the organization of the brain's white matter or \emph{connectome}. Controllability \cite{kailath1980linear} and synchronizability \cite{barahona2002synchronization} separately predict the brain's ability to transition to nearby \emph{versus} distant states, or to maintain a single state characterized by a stable temporal dynamic. While mathematically, there are no known correspondences between these two constructs, we uncover evidence that the brain optimizes the former (controllability, to both near and distant states) at the expense of the latter (synchronizability). Perhaps even more notable, this optimization occurs during development in youth aged 8 to 22 years, and individual differences in control architecture of white matter are correlated with individual differences in cognitive performance. We use forward-modeling computational approaches \cite{Avena2014} to identify constrained evolutionary trajectories, providing evidence that network control is a key mechanism in development \cite{gu2015controllability}. 

In considering our analyses and results, it is worth mentioning some considerations related to biophysical relevance. Specifically, one might ask how these constructs of network controllability relate to constructs of control in the brain? Here we use a simple linear model of brain network dynamics to estimate the statistics of average and modal controllability. Importantly, the same linear model has been used previously to understand both the relationship between fMRI BOLD data and the structural architecture of white matter tracts estimated from diffusion imaging data \cite{Honey2009}, and how the synaptic connections between neurons determine the repertoire of spatial patterns displayed in spontaneous activity \cite{Galan2008}. While this simple linear model is useful in understanding first-order dynamics in networked systems, such linearization of the dynamics does constrain the model's predictive power to short time scales and to states in the immediate vicinity of the operating point \cite{gu2015controllability}.

Further work has built on the simple linear model of brain dynamics that we study here by seeking to understand how underlying anatomical structure constrains the potential impact of control energy localized to single nodes \cite{gu2015controllability,muldoon2016stimulation} or spanning multiple nodes \cite{betzel2016optimally, gu2017optimal} (so-called ``multi-point'' control). As described in a recent review \cite{tang2017control}, the input control energy can be thought of as derived from any process that changes the activity of a single (or multiple) region(s). Common examples include but are not limited to the activation elicited by external stimuli, and changes in electrical activity elicited by brain stimulation. Previous work has shown that average versus modal control is more effectively enacted by default mode versus executive regions, and thus that there is a natural mapping between the control type and the function performed by different brain areas at the large scale \cite{gu2015controllability}. Moreover, there is evidence that while a region may -- on average -- show a preference for a certain control strategy, it may also play a different control role depending on the task context: i.e., which initial state of the brain must be moved to which final state \cite{betzel2016optimally,gu2017optimal}. In the context of this particular study, these approaches allow us to examine the local controllability of each node and the structure of dynamics supported on it using aggregate statistics, for the goal of comparing the relative control strengths of different regions.

Our findings regarding the anatomical localization of control profiles are particularly interesting when considered against the backdrop of prior empirical work describing the neurophysiological dynamics supporting cognitive control. For example, high modal controllers -- predominantly found in executive areas \cite{gu2015controllability} -- are predicted to control dynamics of the brain into distant, difficult-to-reach states. These inferences are consistent with, and provide novel structurally-based neural mechanisms for, the observed empirical function of cognitive control areas \cite{Wallis2015}. Specifically, cognitive control areas are thought to drive or constrain neurophysiological dynamics over distributed neural circuits using transient modulations \cite{chai2017evolution}, consistent with the role of modal controllers \cite{gu2015controllability}. Conversely, high average controllers -- predominantly found along the medial wall \cite{gu2015controllability} -- are predicted to control the brain's intrinsic dynamics towards nearby states, potentially explaining the competitive relationships observed between cognitive control areas and medial portions of the default mode system \cite{Fornito2012}. More generally, the role of structural connectivity underpinning these large-scale coordinated processes is not commonly accounted for in current computational models of cognitive control \cite{Botvinick2014}. It will be important to understand how these structural drivers constrain high-frequency activity in both health and in disorders accompanied by executive deficits, particularly because such an understanding could inform novel interventions with the network biology.

The theoretical links between network control and executive function are particularly intriguing in light of our observations that brains predicted to switch easily to nearby mental states are also predicted to switch easily to distant mental states. Given that the brain regions high in average controllability are different from those high in modal controllability, this positive relationship was unexpected; one might intuitively assume that a brain with high performance on one type of control strategy would display low performance on another. Indeed, in many computational studies of brain network architecture, the common finding is that a network optimized for one type of structure (such as local clustering) will not display another type of structure (such as modular organization) \cite{Klimm2014}. Our results suggest that individual differences in network control are correlated. This may partly explain the fact that different types of cognitive abilities tend to be highly correlated: individuals who are good at one type of cognitive task tend to be good at other cognitive tasks \cite{Miyake2000}.

Beyond their implications for individual differences in cognition, our results also shed important light on brain development. Specifically, our approach reveals the emergence of regional super-controllers as youth between the ages of 8 and 22 years mature. These findings suggest that there is a fundamental change in graph architecture that enables specialization of regional function. Indeed, structural changes in white matter microstructure within specific brain areas have previously been linked to functional specialization, largely in terms of the computations that are being performed \cite{Jarbo2015}. The super-controllers we identify here broaden these findings by suggesting that large-scale changes in network architecture support the emergence of regions specialized for different types of control strategies and different length-scales of coordination.  Critically, average super-controllers are located in a broad swath of frontal-parietal cortex, which is well-known to support the emergence of executive functions and the acquisition of new behaviors \cite{Chrysikou2011}. Modal super-controllers are located in prefrontal areas that play a critical role in the emergence of cognitive control \cite{Seghete2013}. Notably, individual differences in cognitive ability -- above and beyond those explained by age -- are driven by relatively stable-controllers in subcortical regions. These results suggest that the relative strength of controllers in subcortical \emph{versus} cortical regions is critical for understanding individual differences in overall cognitive function, a notion that is supported by the functional segregation of these areas in healthy adults \cite{Cerliani2015}. Lastly, we note that the patterns in white matter architecture in children have a greater potential to increasingly support distant (difficult) brain state transitions, whereas the potential to support nearby (easy) brain state transitions remains constant over development. 

Finally, it is worth offering a few speculations regarding potential future directions for this type of work. Our observation that brain controllability increases during neurodevelopment suggests the existence of an optimization process that maximizes the human brain's ability to transition among mental states while minimizing our vulnerability to being fixed in a single state. If this suggestion is true, then what specific neurophysiological dynamics are enhanced by this increased controllability? What behavioral phenotypes would these optimizations support? Answers to these and related questions will require new directions of empirical research seeking to bridge the neurophysiological drivers of skill acquisition with the control architectures that support them \cite{gu2015controllability,pasqualetti2014controllability}. Such studies might shed light on the question of whether structural changes enable the learning of new behaviors, or whether learning itself alters white matter architecture such that the control energy required for a task decreases as a youth matures. These questions would benefit from longitudinal empirical studies and provide a step towards characterizaton of healthy neurodevelopment.

Lastly, our mechanistic modeling efforts sought to investige the rearrangement of network topology through edge swaps in the human brain network. Interestingly, this approach mimics an aspect of neural plasticity and reorganization that may naturally occcur through adolescent development \cite{Mitra2012}. Future work could expand on this model to take into account spatial constraints on brain network architecture \cite{bassett2010efficient}, and implement addition and deletion of edges tracking the known trajectories of growth or pruning processes. While our findings support the notion that optimization of controllability is a mechanism in development, more detailed biophysical investigation is needed for a complete characterization.

\section*{Methods}
\subsection{Subjects}

All data were acquired from the Philadelphia Neurodevelopmental Cohort (PNC), a large community-based study of brain development. This resource is publicly available through the Database of Genotypes and Phenotypes. This study includes 882 subjects between 8--22 years old (mean age=15.06, SD=3.15, see Supplementary Fig. 2; 389 males, 493 females), each of whom provided their informed consent according to the Institutional Review Board of the University of Pennsylvania who approved all study protocols. These subjects had no gross radiological abnormalities that distorted brain anatomy, no history of inpatient psychiatric hospitalization, no use of psychotropic medications at the time of scanning, and no medical disorders that could impact brain function. Each of the 882 included subjects also passed both manual and automated quality-assessment protocols for DTI \cite{Roalf2016903} and T1-weighted structural imaging \cite{Vandekar14012015}, and had low in-scanner head motion (less than 2mm mean relative displacement between \textit{b}=0 volumes). We acknowledge that using only data of high quality does not overcome all of the inherent limitations of deterministic or probabilistic tractography algorithms \cite{Maier-Hein084137}, but reducing noise in the diffusion weighted data results in better tract estimation and reduced false positives, as recently documented \cite{Maier-Hein084137}.

\subsection{Diffusion tensor imaging}

Diffusion tensor imaging (DTI) data and all other MRI data were acquired on the same 3 T Siemens Tim Trio whole-body scanner and 32-channel head coil at the Hospital of the University of Pennsylvania. DTI scans were obtained using a twice-refocused spin-echo (TRSE) single-shot EPI sequence (TR = 8100ms, TE = 82ms, FOV = 240mm$^2/240$mm$^2$; Matrix = RL: 128/AP:128/Slices:70, in-plane resolution (x \& y) 1.875 mm$^2$; slice thickness = 2mm, gap = 0; FlipAngle = $90^{\circ}/180^{\circ}/180^{\circ}$, volumes = 71, GRAPPA factor = 3, bandwidth = 2170 Hz/pixel, PE direction = AP). The sequence employs a four-lobed diffusion encoding gradient scheme combined with a 90-180-180 spin-echo sequence designed to minimize eddy-current artifacts. The complete sequence consisted of 64 diffusion-weighted directions with $b = 1000$s/mm$^2$ and 7 interspersed scans where $b = 0$s/mm$^2$. Scan time was approximately 11 min. The imaging volume was prescribed in axial orientation covering the entire cerebrum with the topmost slice just superior to the apex of the brain \cite{Roalf2016903}.

\subsection{Cognitive testing}

Cognitive scores were measured using tests from the Penn Computerized Neurocognitive Battery, from which a bifactor analysis revealed a summary efficiency score that we utilized as a measure of subject cognitive performance \cite{ref2014-36496-001}. We used cognitive scores for 880 subjects from the original sample, which passed the quality control measures for cognitive testing.

\subsection{Connectome construction}

Structural connectivity was estimated using 64-direction DTI data. The diffusion tensor was estimated and deterministic whole-brain fiber tracking was implemented in DSI Studio using a modified FACT algorithm, with exactly 1,000,000 streamlines initiated per subject after removing all streamlines with length less than 10mm \cite{gu2015controllability}. A 234-region parcellation \cite{10.1371/journal.pone.0048121} was constructed from the T1 image using FreeSurfer. Parcels were dilated by 4mm to extend regions into white matter, and registered to the first non-weighted (\textit{b}=0) volume using an affine transform. Edge weights $A_{ij}$ in the adjacency matrix were defined by the number of streamlines connecting each pair of nodes end-to-end. All analyses were replicated using an alternative edge weight definition, where weights are equal to the number of streamlines connecting each node pair divided by the total volume of the node pair, as well as using probabilistic fiber tracking methods (see following section). The schematic for structural connectome construction is depicted in Fig. 1a.

Brain regions within the 234-region parcellation can be assigned to anatomical and cognitive systems. We use this assignment to identify 14 subcortical brain regions in both the left and right hemispheres: the thalamus proper, caudate, putamen, pallidum, accumbens area, hippocampus and amygdala.

\subsection{Network controllability}
A networked system can be represented by the graph $\mc G = (\mc V, \mc E)$, where $\mc V$ and $\mc E$ are the vertex and edge sets, respectively. Let $a_{ij}$ be the weight associated with the edge $(i,j) \in \mc E$, and define the \emph{weighted adjacency matrix} of $\mc G$ as $A = [a_{ij}]$, where $a_{ij} = 0$ whenever $(i,j) \not\in \mc E$. We associate a real value (\emph{state}) with each node, collect the node states into a vector (\emph{network state}), and define the map $\map{x}{\mathbb{N}_{\ge 0}}{\mathbb{R}^n}$ to describe the evolution (\emph{network dynamics}) of the network state over time. 

In our case, $\mathbf{A} \in \real^{N \times N}$ is a symmetric and weighted adjacency matrix whose elements indicate the number of white matter streamlines connecting two different brain regions --- denoted here as $i$ and $j$. An underlying assumption of this approach is that the number of streamlines is proportional to the strength of structural connectivity.

\subsection{Dynamical model}

The equation of state that we use is based on extensive prior work demonstrating its utility in predicting resting state functional connectivity \cite{Honey2009} and in providing similar brain dynamics to more complicated models \cite{Galan2008}. Although neural activity evolves through neural circuits as a collection of \emph{nonlinear} dynamic processes, these prior studies have demonstrated that a significant amount of variance in neural dynamics as measured by fMRI can be predicted from simplified \emph{linear} models. 

Based on this literature, we employ a simplified noise-free linear discrete-time and time-invariant network model \cite{gu2015controllability}: \begin{equation}\label{eq: linear network}
  \mathbf{x} (t+1) = \mathbf{A} \mathbf{x}(t) + \mathbf{B}_{\mc K} \mathbf{u}_{\mc K} (t) ,
\end{equation}
where $\map{\mathbf{x}}{\real_{\ge 0}}{\real^N}$ describes the state (i.e., a measure of the electrical charge, oxygen level, or firing rate) of brain regions over time, and $\mathbf{A} \in \real^{N \times N}$ is the structural connectome described in the previous section. Hence the size of the vector $\mathbf{x}$ is given by the number of brain regions in the parcellation (e.g., 234 under the Lausanne parcellation, see Methods section, in the subsection entitled ``Connectome construction''), and the value of $\mathbf{x}$ describes the brain activity of that region, such as the magnitude of the BOLD signal.

The diagonal elements of the matrix $\mathbf{A}$ satisfy $A_{ii}=0$. Note that to assure Schur stability, we divide the matrix by $1+\xi_0(\mathbf{A})$, where $\xi_0(\mathbf{A})$ is the largest singular value of $\mathbf{A}$. The input matrix $\mathbf{B}_{\mc K}$ identifies the control points $\mc K$ in the brain, where $\mc K = \{k_1, \dots, k_m \}$ and \begin{align}\label{eq: B}
  B_{\mc K} =
  \begin{bmatrix}
    e_{k_1} & \cdots & e_{k_m}
  \end{bmatrix},
\end{align}
and $e_i$ denotes the $i$-th canonical vector of dimension $N$. The input $\map{\mathbf{u}_{\mc K}}{\real_{\ge  0}}{\real^m}$ denotes the control strategy.

We study the \emph{controllability} of this dynamical system, which refers to the possibility of driving the state of the system to a specific target state by means of an external control input \cite{rek-ych-skn:63_2}. Classic results in control theory ensure that controllability of the network \eqref{eq: linear network} from the set of network nodes $\mc K$ is equivalent to the controllability Gramian $\mathbf{W}_{\mc K}$ being invertible, where 
\begin{equation}
  \mathbf{W}_{\mathcal{K}} = \sum_{\tau =0}^{\infty}\mathbf{A}^\tau
  \mathbf{B}_{\mathcal{K}}\mathbf{B}_{\mathcal{K}}^\transpose \mathbf{A}^\tau .
\end{equation}
Consistent with \cite{gu2015controllability}, we use this framework to choose control nodes one at a time, and thus the input matrix $\mathbf{B}_{\mc K}$ in fact reduces to a one-dimensional vector, e.g., $\mathbf{B}_{\mc K}=\begin{pmatrix}1& 0& 0& ... \end{pmatrix}^T$ when the first brain region is the control node. In this case, $\mc K$ simply describes this control node, i.e. the controllability Gramian can be indexed by the $i$-th control node that it describes: $\mathbf{W}_{i}$.

While the brain certainly displays non-linear activity, modelling of brain activity in large-scale regional networks shows that the linear approximation provides fair explanatory power of resting state fMRI BOLD data \cite{Honey2009}. Further, studies of this controllability framework using non-linear oscillators connected with coupling constants estimated from large-scale white matter structural connections shows a good overlap with the linear approximation \cite{muldoon2016stimulation}. While the model we employ is a discrete-time system, this controllability Gramian is statistically similar to that obtained in a continuous-time system \cite{gu2015controllability}, through the comparison of simulations run using MATLAB's \textit{lyapunov} function.

\subsection{Controllability metrics}

Within this controllability framework, we study two different control strategies that describe the ability to move the network into different states defined as patterns of regional activity (Fig. 1b). Average controllability describes the ease of transition to many states nearby on an energy landscape, while modal controllability describes the ease of transition to a state distant on this landscape.

Average controllability of a network equals the average input energy from a set of control nodes and over all possible target states. As a known result, average input energy is proportional to $\text{Trace}( \mathbf{W}_{\mathcal{K}} ^{-1})$, the trace of the inverse of the controllability Gramian. Instead and consistent with \cite{gu2015controllability}, we adopt $\text{Trace}( \mathbf{W}_{\mathcal{K}} )$ as a measure of average controllability for two main reasons: first, $\text{Trace}( \mathbf{W}_{\mathcal{K}} ^{-1})$ and $\text{Trace}( \mathbf{W}_{\mathcal{K}} )$ satisfy a relation of inverse proportionality, so that the information obtained from the two metrics are correlated with one another and, second, $ \mathbf{W}_{\mathcal{K}} $ is typically very ill-conditioned even for coarse network resolutions, so that $\text{Trace}( \mathbf{W}_{\mathcal{K}} ^{-1})$ cannot be accurately computed even for small brain networks. It should be noted that $\text{Trace}( \mathbf{W}_{\mathcal{K}} )$ encodes a well-defined control metric, namely the energy of the network impulse response or, equivalently, the network $H_2$ norm \cite{kailath1980linear}. As discussed above, when a brain region $i$ forms a control node, the resulting Gramian can be indexed as $\mathbf{W}_{i}$, in order to compute the regional average controllability.

Modal controllability refers to the ability of a node to control each
  evolutionary mode of a dynamical network, and can
  be used to identify states that are difficult to control from a set
  of control nodes. Modal controllability is computed from the
  eigenvector matrix $V = [v_{ij}]$ of the network adjacency matrix
  $\mathbf{A}$. By extension from the PBH test \cite{kailath1980linear}, if the
  entry $v_{ij}$ is small, then the $j$-th mode is poorly controllable
  from node $i$. Following \cite{pasqualetti2014controllability}, we define $\phi_i =
  \sum_{j} (1 - \xi_j^2 (\mathbf{A})) v_{ij}^2$ as a scaled measure of
  the controllability of all $N$ modes $\xi_0 (\mathbf{A}),\dots, \xi_{N-1}
  (\mathbf{A})$ from the brain region $i$ -- allowing the computation of regional modal controllability. Regions with high modal
  controllability are able to control all the dynamic modes of the
  network, and hence to drive the dynamics towards
  hard-to-reach configurations.
  
  The mean average controllability over the whole brain network is then the mean over all regional average controllability values, and similarly for mean modal controllability as the mean over all regional modal controllability values. The availability of scripts to calculate these metrics is given at the end of the Methods section, in the subsection entitled ``Data Availability''.

\subsection{Discrete transitions and centralized \emph{versus} distributed control}
Here, we provide some discussion of the evidence from the literature supporting the notion that the brain may make discrete transitions between large-scale states. First, we note that in our current model, a brain state is a pattern of activity across 234 cortical and subcortical areas; thus, a discrete state transition can be constituted by a change in the activity level of a single area. Thus, when we discuss state transitions, not all states must be far from one another -- in fact, many are ``close'' as defined by a Euclidean distance (or L2 norm or similar) between state vectors. Nevertheless, whether states are near or far, it is interesting to review the evidence that discrete brain states (and non-gradual state transitions) can be identified in large-scale neuroimaging data.

Perhaps the simplest illustration of discrete transitions in brain state dynamics is that observed during bistable perception, where fMRI activity is well-fit by a pairwise maximum entropy model with characteristic basin states and estimable transition rates between them \cite{watanabe2014energy}. This and similar work builds on a prior studies demonstrating that neural activity during multistable behaviours can be described as a series of periods in which the system dwells within an attractor and transitions between different attractors on the underlying energy landscape \cite{deco2009stochastic}. Yet, bi- or multi-stable perception is arguably not a common cognitive process experienced in everyday life; therefore, the stronger evidence for the importance of discrete brain state transitions comes from observations made about the brain's intrinsic dynamics, the baseline from which task-based processes are drawn. Indeed, careful work suggests that the brain's intrinsic dynamics as measured by fMRI are also well-fit by a pairwise maximum entropy model with distinct basin states and well-parameterized transition rates between them \cite{watanabe2013pairwise}. Moreover, in data-driven work independent of maximum entropy model assumptions, evidence from multiband resting state fMRI suggests that transitions from low-to-high efficiency states are quite sudden (approximately discrete) and the transitions from high-to-low efficiency states are quite gradual (approximately continuous), a variation in dynamics that is argued to achieve a balance between efficient information-processing and metabolic expenditure \cite{zalesky2014time}. Evidence for longer-time scale state transitions comes from longitudinal imaging experiments that demonstrate that resting state functional connectivity can be separated into two distinct meta-states, and that these states are related to differences in self-reported attention \cite{shine2016temporal}.

We further examine ideas of centralized \emph{versus} distributed control. First, we note that while we study the control capabilities of single brain regions, this does not preclude these regions enacting control in groups. Our motivation in this study is to examine the control capabilities predicted from a single region, as important initial means of gaining fundamental understanding about the system. In other work, we have extended these methods to use the same LTI model to address questions of distributed or multi-point control \cite{betzel2016optimally,gu2017optimal}; in other words, the use of the LTI model does not preclude a study of distributed control. While such a study  is beyond the scope of the current work, it might be a useful direction for future efforts.

Nonetheless, it is interesting to review the evidence for centralized control in cognitive processes provided by neuroimaging data. In considering this topic, it is useful to distinguish between external control, which is enacted on the system from the outside, and internal control, which is a feature of the system itself. In the brain, internal control processes include phenomena as conceptually diverse as homeostasis, which refers to processes that maintain equilibrium of dynamics \cite{nelson2008strength}, and cognitive control, which refers to processes that exert top-down influence to drive the system between various dynamical states \cite{botvinick2015motivation}. Focusing solely on cognitive control, we note that historical models explained the production of decisions based on a given set of inputs using the perceptron \cite{rosenblatt1958perceptron}, a simple artificial neural network \cite{bishop1995neural}. The complexity of the connection architecture in this model was thought to support a complexity of brain dynamics, such as the separation of parallel neural processes and distributed neural representations propounded by the parallel distributed processing (PDP) model \cite{rumelhart1986parallel}. This and related computational models emphasize the role of specific brain areas in cognitive control, including prefrontal cortex, anterior cingulate, parietal cortex, and brainstem \cite{braver2003neural}, which were later referred to as the primary control system of the brain \cite{dosenbach2007distinct}. Yet, some argue that control in the brain is not localized to small regions or modules, but is instead very broadly distributed, enabling versatility in both information transfer and executive control \cite{Eisenreich077685}. Recent data have shed additional light on this controversy, and have moreover broadened the cognitive processes of interest (and by extension the applicability of these models) beyond cognitive control. Specifically, considering rest and three distinct tasks requiring semantic judgments, mental rotation, and visual coherence, Gratton and colleagues provide evidence for two independent factors that modify brain networks to meet task goals \cite{gratton2016evidence}: (i) regions activated in a task (consistent with, for example, \cite{cole2016activity}) and, (ii) regions that serve as connector hubs for transferring information across systems (consistent with, for example, \cite{bertolero2015modular}). Regions that shared both features, so-called ``activated connector'' regions exhibited attributes consistent with a role in enacting task control. These data not only pinpoint specific centralized regions involved in control, but also suggest that the constitution of those regions may depend on the task at hand. Such a suggestion is also consistent with our recent computational study of optimal brain state trajectories from a state in which the default mode is active to a state in which primary visual, auditory, and sensorimotor cortices are active \cite{gu2017optimal}. In this study, we find that the temporo-parietal junction is consistently identified as an optimal controller across all of these state transitions, but there are also task-specific controllers that differ according to the anatomy of the target state. In sum, current literature supports the notion that there may be a centralized control system for cognitive control \cite{braver2003neural}, but that other sorts of brain state transitions might capitalize on distributed strategies that are constrained by the anatomy of the initial and target states \cite{gu2017optimal}.
\subsection{Network synchronizability}
Synchronizability measures the ability of a network to persist in a single synchronous state $\mathbf{s} (t)$, i.e. $\mathbf{x}_1 (t) = ... = \mathbf{x}_n (t+1) = \mathbf{s} (t)$ (see Fig.~2a in main text). The master stability function (MSF) allows analysis of the stability of this synchronous state without detailed specification of the properties of the dynamical units \cite{PhysRevLett.80.2109}. Within this framework, linear stability depends on the positive eigenvalues  $\{\lambda_{i}\}, i=1, ... ,N-1$ of the Laplacian matrix $\mathbf{L}$ defined by $L_{ij}=\delta_{ij}\sum_{k}A_{ik}-A_{ij}$.

The condition for stability depends on the shape of the MSF and whether these eigenvalues fall into the region of stability. Hence we can use the normalized spread of the eigenvalues to quantify how synchronizable the network will generally be \cite{Nishikawa08062010}. We therefore quantify network synchronizability as

 \begin{equation}
 \frac{1}{\sigma^2}=\frac{d^2(N-1)}{\sum_{i=1}^{N-1}|\lambda_i-\bar{\lambda}|^2}\textrm{,\quad where } \bar{\lambda}:=\frac{1}{N-1}\sum_{i=1}^{N-1}\lambda_i
 \label{eq:sync}
 \end{equation}
and $d:=\frac{1}{N}\sum_{i}\sum_{j\neq i}A_{ij}$, the average coupling strength per node, which normalizes for the overall network strength. To illustrate that synchronizability is driven much more by variation in the eigenspectrum (denominator of Eq. \ref{eq:sync}) than by differences in connection density (numerator), we plot separately the numerator and denominator with age, see Supplementary Fig. 1.

We have plotted a typical example of a MSF for a network of oscillators schematically in Fig.~2a; however, specific details will depend on the dynamics on individual nodes and the connectivity between them. The shape of the MSF for various families of dynamical systems is typically convex for generic oscillator systems, including chaotic oscillators that have stable limit cycles \cite{PhysRevE.80.036204}.

\subsection{Network statistics and curve-fitting}

To assess the statistical significance of our results, we constructed non-parametric permutation-based null models.  Specifically, the null models in Fig.~1 retained the same regions as the real network but permuted edge weights uniformly at random within the constraints of preserving degree and strength, respectively. To preserve degree we simply permuted non-zero weights within a network, and to preserve strength we used the function \textit{null\_model\_und\_sign} from the Brain Connectivity Toolbox that permutes edge weights to approximately preserve the strength of each node.

Pearson correlations were predominantly used except where the data distribution was markedly skewed, in which case Spearman correlations were used instead (regional modal controllability and cognitive performance). Regional controllability values were the mean controllability values over all individuals: 190 subjects aged 18 and above in Fig.~1, and all 882 subjects in Fig.~3. To test for the regional significance of correlation with age in Fig.~3, a false discovery rate correction for multiple comparisons was used with $q=0.05$. 

Confidence intervals in the plots of Fig. 2c, d, and 3c were computed in the software R using the \textit{visreg} library, which shows 95\% confidence intervals in grey around the fitted lines. The non-linear fits in Fig. 2c-d were made using a generalized additive model, which is  a generalized linear model where the linear predictor is given by penalized regression splines of the variable plus conventional parametric components of the linear predictors, e.g.
\begin{multline}
\textrm{Mean average controllability}=\textrm{intercept}+ \textrm{spline (age)}
\\+ \textrm{sex} +\textrm{handedness} +\textrm{brain volume}+\textrm{head motion}
\end{multline}
These fits were calculated in the software R using the \textit{mgcv} library, which has previously been used to describe both linear and nonlinear developmental effects in the PNC dataset \cite{Vandekar14012015}.

Curve fitting was done using the Curve Fitting Toolbox in MATLAB. We chose exponential fits for all data and optimization trajectories as three parameters in each case produced good fits. However, the Pareto optimization trajectories are not really exponential, i.e. taking the log of one of the variables does not make the relationship linear. Hence we simply left all plots in their original axes and used exponential fits. 

\subsection{Pareto-optimization parameters}

The trajectories traced out by Pareto-optimization can be very constrained in the paths they delineate and especially in the forward direction of mean average controllability and mean modal controllability. We always ran 100 parallel computations each with their own random edge swaps, and in this direction of forward optimization for controllability all the curves followed the same path. Curve-fitting was done only on trajectories in this direction, for which we simply picked one trajectory out of the 100 similar ones. 

Termination of the Pareto-optimization process was done after 1500 evolutionary steps, by which time the controllability metrics showed comparatively small changes from one step to the next. The mean absolute value of changes in controllability metrics for the last 500 steps were below 1\% of the total change in either mean average controllability or mean modal controllability, for the average subject.

Trajectories in the synchronizability cross-sections and backward direction showed greater variability among the parallel simulations. In the backwards direction, after a smooth decrease in controllability for many steps, some curves began to turn around or display erratic jumps, see Supplementary Fig. 3 for trajectories chosen at random from each subject. These backward trajectories were truncated when the gradient in the controllability plane (change of mean modal controllability over change in mean average controllability) became negative. We then retained the longest trajectory (visualized in Fig.~4), although in most cases there was little loss of overall trajectory length. 

\subsection*{Generation of network models}

We take a few of the most common graph models from the general literature, as well as more specifically from the literature postulating models of the topology observed in human brain networks \cite{Klimm2014}. Below, we briefly describe the 8 graph models that we chose as the basis for the analysis presented in the Results section, specifically in the subsection entitled ``Controlling for Modularity''. Additional details for these models can also be found in our recent publication \cite{sizemore2017classification}.

\noindent  1. \textit{(WRG) Weighted Random Graph model}: Arguably the most fundamental, this graph model is a weighted version of the canonical Erd\H{o}s-R\'{e}nyi model. For all pairs of $N$ nodes, we modeled the weight of the edge by a geometric distribution with probability of success $p$, the desired edge density of the graph. Each edge weight was assigned the number of successes before the first failure \cite{sizemore2017classification}.

\noindent 2. \textit{(RL) Ring Lattice model}: In contrast to the random nature of the WRG, the ring lattice model is one with strict order. We arranged $N$ nodes on the perimeter of a regular polygon, each with degree $k$, determined by the desired edge density. We then connected each node to the $\frac{k}{2}$ nodes directly before and after it in the sequence of nodes on the polygon. Edge weights were assigned the inverse of the path length between the two nodes, determined by traversing the perimeter of the polygon \cite{sizemore2017classification}.

\noindent 3. \textit{(WS) Watts-Strogatz model}: A model that bridges both the order of the RL and the disorder of the WRG, the Watts-Strogatz graph model is a ring lattice model in which edges are rewired uniformly at random to create a small-world network. Following \cite{sizemore2017classification}, we chose the probability $r$ of rewiring a given edge to maximize the small-world propensity \cite{muldoon2016small}.

\noindent 4. \textit{(MD2) Modular Network with 2 communities model}: While the previous models can display some local clustering structure, they lack meso-scale organization in the form of modules or communities. In contrast, the Modular Network with 2 communities model is a graph of $N$ nodes and $K$ edges placed so as to form 2 communities. Pairs of nodes \textit{within} communities are connected with edge density 0.8, and pairs of nodes \textit{between} communities (where one node in the pair is in one community and the other node in the pair is in a different community) are connected to fulfill the desired total edge density $p$. We assigned weights to existing edges by considering a geometric distribution with probability of success $p$ if the nodes were in the same module and $1-p$ if the nodes were in different modules. Each edge weight was assigned the number of successes before the first failure \cite{sizemore2017classification}.

\noindent 5. \textit{(MD4) Modular Network with 4 communities model}: This model is generated in a manner identical to that used in the MD2 graph model, with the exception that MD4 has 4 communities.

\noindent 6. \textit{(MD8) Modular Network with 8 communities model}: This model is generated in a manner identical to that used in the MD2 graph model, with the exception that MD8 has 8 communities.

\noindent 7. \textit{(RG) Random Geometric model}: In contrast to most of the previous graph models that were agnostic to any embedding space, the Random Geometric model contains $N$ nodes, chosen randomly from a unit cube, and edges whose weights were equal to the inverse of the Euclidian distance between two nodes. We kept only the $K$ shortest edges, in order to maintain the desired edge density $p$  \cite{sizemore2017classification}.

\noindent 8. \textit{(BA) Barab\'{a}si-Albert model}: In our final graph model, we used software from \cite{Klimm2014} to generate a typical BA model -- a scale-free network that exhibits preferential attachment to existing nodes of high degree -- with $N$ nodes and $K$ edges. Each edge weight was assigned the average degree of the two nodes it connected.

In all of the graph models described above, two parameters are fixed \emph{a priori}: the number of nodes $N$ in the network, and the number of edges $K$ in the network. Several of the graph models are defined only for a cardinality that is a power of 2; to include these models, and also speed computation time, we chose the number of nodes to be 128. Further, we chose the number of edges to produce network densities that were consistent with those observed empirically in large-scale human brain graphs. To maximize generalizability of our findings to other studies, we chose an independent data set of 19-minute multiband diffusion spectrum imaging from 30 healthy adult individuals \cite{betzel2016optimally}, with an average edge density of 0.2919. For each network size, we generated 100 instantiations of each of the 8 graph models described above.

All of the 8 graph models described above were weighted graph models \cite{sizemore2017classification}. While it is important to study weighted (as opposed to binary) graph models to benchmark our metrics that are currently being applied to real-world weighted graphs, comparisons across models are confounded by the fact that each model can have a very different edge weight distribution. Here, we sought to disentangle the impact of graph model from the impact of edge weight distribution on network controllability statistics. Accordingly, we therefore developed a pipeline to reweight all of the graph models fairly, and with a fixed edge weight distribution.

We began by adding random noise on the order of $10^{-7}$ to all edge weights in all network models; this process ensures the uniqueness of each edge weight, while maintaining the relative weight magnitudes. Next, we sorted edges by weight, and then replaced each edge with corresponding ordered values pulled from a specific edge weight distribution of interest, of which we defined three. The first was a Gaussian distribution with a mean of 0.5 and a standard deviation of 0.12, ubiquitously found in real-world networks. The second edge weight distribution of interest was taken from \cite{betzel2016optimally} to closely model empirical weighting distributions in large-scale human brain structural networks estimated from diffusion imaging tractography. Specifically, this distribution was based on streamline counts, normalized by the geometric mean of regional volumes, as we use in the current study. Importantly, the reweighting scheme we describe here allowed us to use the same edge weighting across all graphs to guarantee that differences in controllability were due to topology and not to other properties of the graphs, like differing edge weights and scaling.

In summary, we study network ensembles for two types of edge weightings: streamline counts and Gaussian weights. Each of these ensembles includes 100 instantiations of each of the 8 graph models. 

\subsection*{Data availability}
Scripts to calculate the controllability metrics can be found at www.danisbassett.com/research-projects.html for public use.

Diffusion tensor imaging data were acquired from the Philadelphia Neurodevelopmental Cohort, a resource publicly available through the Database of Genotypes and Phenotypes.

~\\
~\\
\paragraph{Acknowledgments}
E.T. and D.S.B. would like to acknowledge support from the John D. and Catherine T. MacArthur Foundation, the Alfred P. Sloan Foundation, the Army Research Laboratory and the Army Research Office through contract numbers W911NF-10-2-0022 and W911NF-14-1-0679, the National Institute of Health (2-R01-DC-009209-11, 1R01HD086888-01, R01-MH107235, R01-MH107703, R01MH109520, 1R01NS099348 and R21-M MH-106799), the Office of Naval Research, and the National Science Foundation (BCS-1441502, CAREER PHY-1554488, BCS-1631550, and CNS-1626008). The content is solely the responsibility of the authors and does not necessarily represent the official views of any of the funding agencies.

\paragraph{Author contributions}
E.T., C.G., D.S.B. and T.D.S. designed the research; E.T. performed the research and analysis; G.B. performed the tractography; S.G. and E.P. wrote code for the controllability metrics and Pareto optimizations, respectively; A.K., D.R. and K.R. preprocessed the DTI data; T.M.M. preprocessed the cognitive performance data; T.D.S., R.C.G. and R.E.G. oversaw the data collection; while E.T., D.S.B., and T.D.S. wrote the paper.

\paragraph{Conflict of Interest}
The authors declare no competing financial interests.

\section{References}
\bibliography{bibfile_elena,braincontrol,bibfile_new_2,bibfile_2,bibfile_new,bibfile_original,bibfile,dynamical_trajectories_v5_discussionOnly_db,dynamical_trajectories_v5_structureFunction_RB,control_sync_DB}
\begin{table*} [ht]
 \centering
 \caption {\textbf{Relationship between mean average controllability and modularity $Q$ for various graph types.} Spearman $\rho$-values and corresponding $p$-values for the correlations between mean average controllability and modularity $Q$. The two weighting schemes used are Gaussian weights and streamline counts, while the eight graph models are the weighted random graph (WRG), the ring lattice (RL), the Watts-Strogatz small-world (WS), the modular graphs (MD2, MD4, MD8), the random geometric (RG), and the Barabasi-Albert preferential attachment (BA) models. The $p$-values stated to be zero are below $10^{-5}$.}
\begin{tabular}{|c|c|c|c|c|c|c|c|c|}
\hline
\textbf{128 nodes} & WRG & RL & WS & MD2 & MD4 & MD8 & RG & BA \\
\hline
\hline
Gaussian & ~ & ~ & ~ & ~ & ~ & ~ & ~ & ~ \\
\hline
$\rho$ & 
 -0.0297 &  -0.1632 &  -0.2148 &  -0.1689 &  -0.2057 &  -0.0476 &  -0.4568  & -0.0509\\
$p$ &     0.7691   & 0.1046  &  0.0321  &  0.0930   & 0.0403 &   0.6374   & 0.0000  &  0.6144
 \\
\hline
Streamline counts & ~ & ~ & ~ & ~ & ~ & ~ & ~ & ~ \\
\hline
$\rho$ &   -0.0376 &  -0.0539  &  0.1424 &  -0.3642 &  -0.1469  & -0.1107  & -0.0962 &  -0.7538 \\
$p$ &    0.7097  &  0.5939 &   0.1575  &  0.0002  &  0.1444 &   0.2724   & 0.3404     &    0.0000 \\
\hline
\end{tabular} \label{tab:spearman2}
\end{table*}
\end{document}